\documentclass[12pt]{iopart}
\expandafter\let\csname equation*\endcsname\relax
\expandafter\let\csname endequation*\endcsname\relax
\usepackage{amsmath}
\usepackage{physics}
\usepackage{appendix}
\usepackage{bbold}
\usepackage{amsfonts}
\usepackage{amssymb}
\usepackage{amsmath}
\usepackage{enumitem}
\usepackage{xcolor}
\usepackage{suffix}
\usepackage{tikz}
\usepackage{quantikz}
\usepackage{pdfpages}
\usepackage{soul}
\usepackage{theorem}
\usepackage{placeins}

\newcommand{\definitionname}{Definition}
\newcommand{\remarkname}{Remark}
\newcommand{\examplename}{Example}
\newenvironment{definition}[1][\definitionname]{\vspace{.5em}\par\noindent\textbf{#1:} }{\par\vspace{.5em}}
\newenvironment{remark}[1][\remarkname]{\vspace{.5em}\par\noindent\textbf{#1:} }{\par\vspace{.5em}}
\newenvironment{example}[1][\examplename]{\vspace{.5em}\par\noindent\textbf{#1:} }{\par\vspace{.5em}}

\newtheorem{property}{Property}[subsection]

\newcommand\todo[1]{\textcolor{red}{\textbf{\Large TODO:#1}}}
\WithSuffix\newcommand\todo*[1]{}

\begin{document}

\title[Elucidating the Physical and Mathematical Properties of the PTM Sequence in QC]{Elucidating the Physical and Mathematical Properties of the Prouhet-Thue-Morse Sequence in Quantum Computing}

\author{Denis Jankovi\'c}
\address{Institut de Physique et Chimie des Mat\'eriaux de Strasbourg UMR-7504 CNRS, Universit\'e de Strasbourg, 23 rue du Loess, Strasbourg, 67000, France}
\address{{Institute of Nanotechnology}, {Karlsruhe Institute of Technology}, Kaiserstra{\ss}e 12, Karlsruhe, 76131, Germany}
\ead{denis.jankovic@ipcms.unistra.fr}
\author{Rémi Pasquier}
\address{Institut de Physique et Chimie des Mat\'eriaux de Strasbourg UMR-7504 CNRS, Universit\'e de Strasbourg, 23 rue du Loess, Strasbourg, 67000 France}
\address{Institute of Theoretical Physics and Regensburg Center for Ultrafast Nanoscopy, University of Regensburg, Regensburg, 93059,
Germany}
\author{Jean-Gabriel Hartmann}
\address{Institut de Physique et Chimie des Mat\'eriaux de Strasbourg UMR-7504 CNRS, Universit\'e de Strasbourg, 23 rue du Loess, Strasbourg, 67000 France}
\address{IPCMS}
\author{Paul-Antoine Hervieux}
\address{Institut de Physique et Chimie des Mat\'eriaux de Strasbourg UMR-7504 CNRS, Universit\'e de Strasbourg, 23 rue du Loess, Strasbourg, 67000 France}
\address{IPCMS}
\vspace{10pt}

\begin{abstract}
This study explores the applications of the Prouhet-Thue-Morse (PTM) sequence in quantum computing, highlighting its mathematical elegance and practical relevance. We demonstrate the critical role of the PTM sequence in quantum error correction, in noise-resistant quantum memories, and in providing insights into quantum chaos. Notably, we demonstrate how the PTM sequence naturally appears in Ising X-X interacting systems, leading to a proposed robust encoding of quantum memories in such systems. Furthermore, connections to number theory, including the Riemann zeta function, bridge quantum computing with pure mathematics. Our findings emphasize the PTM sequence's importance in understanding the mathematical structure of quantum computing systems and the development of the full potential of quantum technologies and invite further interdisciplinary research.
\end{abstract}

\submitto{\jpa}
%
\maketitle
%
%

\section{Introduction}

The intersection of quantum computing and mathematics opens a fascinating frontier for research, with potential implications for both theoretical advancements and practical applications. Among mathematical constructs, the Prouhet-Thue-Morse (PTM) sequence stands out because of its surprisingly rich structure and multifaceted utility. Originally discovered in the 19th century \cite{Prouhet1851} and renowned for its unique properties, the PTM sequence has garnered attention across various scientific disciplines \cite{Allouche1999}.

Quantum computing, leveraging the principles of quantum mechanics, potentially offers unprecedented computational power and capabilities. However, this potential is often hindered by issues such as quantum error correction, noise resilience, and the intricate dynamics of quantum systems. This study seeks to bridge the gap between the abstract mathematics of the PTM sequence and the practical challenges faced in quantum computing by demonstrating its natural emergence within quantum frameworks.

Intriguingly, we aim to show that it appears very naturally within the Hilbert spaces of quantum computing systems, suggesting profound links and practical applications that are yet to be fully explored. By investigating this manifestation of the PTM sequence, we aim to uncover how its unique properties can address critical issues such as error correction and noise resilience. We show how the PTM sequence's inherent symmetrical structure and fractal nature present opportunities for developing robust quantum error correction codes, thereby enhancing the reliability and efficiency of quantum computations. Additionally, we demonstrate how the PTM sequence appears naturally and easily in X-X Ising systems, further showcasing its significance. Furthermore, the sequence's integration into quantum memory systems and its role in understanding quantum chaos through constructs like the Walsh-Hadamard transform and the quantum baker's map illustrate its broader relevance.

Moreover, this study explores the deep connections between the PTM sequence and number theory, particularly its relation to the Riemann zeta function and, more generally, Dirichlet series. These connections not only enrich our understanding of the mathematical foundations underlying quantum computing but also pave the way for interdisciplinary research that could unlock new computational methods and insights.

In summary, this study aims to establish a comprehensive link between the Prouhet-Thue-Morse sequence and quantum computing, highlighting how the sequence naturally emerges within the computational space of quantum systems. By emphasizing the sequence's potential to enhance error correction, noise resilience, and our understanding of quantum dynamics, we endeavor to illustrate how bridging this gap can contribute to the evolution of quantum technologies and the realization of their full potential.

\subsection{The Prouhet-Thue-Morse sequence}
The Prouhet-Thue-Morse (PTM) sequence is a binary sequen3ce with a rich mathematical structure and wide-ranging applications in physics, from quasicrystals to quantum mechanics \cite{Allouche1999,PhysRevB.43.1034}. 

\begin{definition}
    Mathematically, the sequence is defined by its $n$-th term $t_n$, which is the sum of the binary digits of $n$ (mod 2), formally given by:
\begin{equation}
    t_n =\left( \sum_{k=0}^{\infty} \left\lfloor \frac{n}{2^k} \right\rfloor\right) \text{  mod }2.
\end{equation}
\end{definition}

\begin{example}
\begin{equation}
    \begin{alignedat}{3}
        (0)_2 &= 0,\quad &&t_0 = 0  \text{  mod }2 &&= 0,\\
        (1)_2 &= 1,\quad &&t_1 = 1  \text{  mod }2 &&= 1,\\
        (2)_2 &= 10,\quad &&t_2 = (1 + 0)  \text{  mod }2 &&= 1,\\
        (3)_2 &= 11,\quad &&t_3 = (1 + 1) \text{  mod }2 &&= 0,\\
        (4)_2 &= 100,\quad &&t_4 = (1 + 0 + 0) \text{  mod }2 &&= 1, ...
    \end{alignedat}
\end{equation}
\end{example}

Historically, the sequence was first discovered by Eugène Prouhet in 1851 \cite{Prouhet1851}, later independently rediscovered by Axel Thue in 1906 \cite{Thue1906}, and again by Marston Morse in the 1920s \cite{MorseRecurrentGO}. Its discovery and subsequent investigations have revealed deep connections to various fields of mathematics and physics, illustrating the sequence's profound universality and versatility.

Beyond this initial, explicit, definition, the PTM sequence can also be constructed through iterative or recursive methods.

\begin{definition}
    The iterative approach involves starting with the initial term \(0\) and then repeatedly appending the binary complement of the sequence so far. Formally, if \(T_i\) represents the sequence at the $i$-th iteration (with $2^i$ elements), then:
\begin{equation}
    T_0 = \{0\}, \qquad T_{i+1} = T_i \, || \, \overline{T_i},
\end{equation}
where \(\overline{T_i}\) denotes the binary complement of \(T_i\), and \(||\) denotes concatenation.
\end{definition}

\begin{example}
\begin{equation}
\begin{aligned}
    T_0 &= \{0\},\\
    T_1 &= \{0\}\, || \, \overline{\{0\}} = \{0\}\, || \, \{1\} = \{01\}, \\
    T_2 &= \{01\} \, || \, \overline{\{01\}} =  \{01\}\, || \, \{10\} = \{0110\},\\
    T_3 &=  \{0110\}\, || \, \{1001\} = \{01101001\}, ...
\end{aligned}
\end{equation}
\end{example}

\begin{definition}
Another method is through recursion, where the \(n\)-th term can be defined in relation to its predecessors, given by:
\begin{equation}
    t_n =\left( t_{\lfloor n/2 \rfloor} + n \right)\mod 2, \quad t_0 = 0.
\end{equation}
\end{definition}
These definitions reflect the \textit{self-similarity} and \textit{fractal} nature of the Prouhet-Thue-Morse sequence, emphasizing its deep mathematical properties and relevance to the study of non-periodic structures in physics \cite{Allouche1999,PhysRevB.43.1034, xiong2022topological, DENG2011360, Matarazzo2010, 6326756}.

\begin{definition}
Another similar recursive formula is:
\begin{equation}
       t_0 = 0, \quad
       t_{2n} = t_n, \quad
       t_{2n+1} = 1 - t_n \equiv \overline{t_n}.
\end{equation}
\end{definition}

Finally, an interesting property of the PTM sequence can be described as follows:
\begin{definition}
We define two sets based on the value of $t_n$ within the first $2^N$ terms of the sequence:
\begin{itemize}
    \item $E(N)$, the set of indices $n$ for which $t_n = 0$, for all $n$ in the range $0 \leq n \leq 2^{N} - 1$.
    \item $O(N)$, the set of indices $n$ for which $t_n = 1$, for all $n$ in the same range.
\end{itemize}
\end{definition}
\begin{property}
\label{pro::multigrade}
For any natural number $N$ and for all integers $k$ such that $0 \leq k < N$:
\begin{equation}
\label{eq:tarryescott}
    \sum_{e \in E(N)} e^k = \sum_{o \in O(N)} o^k.
\end{equation}
i.e. the sum of the $k$-th powers of the elements in $E(N)$ equals the sum of the $k$-th powers of the elements in $O(N)$.
\end{property}

\begin{example} For $N=3$, one has
\begin{gather}
\definecolor{Blue}{rgb}{0.29,0.37,0.89}
\definecolor{Red}{rgb}{0.82,0.01,0.11}
    \begin{aligned}
    T_3 = \{&\textcolor{Blue}{0},\, \textcolor{Red}{1},\, \textcolor{Red}{1},\, \textcolor{Blue}{0},\, \textcolor{Red}{1},\, \textcolor{Blue}{0},\, \textcolor{Blue}{0},\, \textcolor{Red}{1}\}\\
    \text{index }\, & \textcolor{Blue}{0},\, \textcolor{Red}{1},\, \textcolor{Red}{2},\, \textcolor{Blue}{3},\, \textcolor{Red}{4},\, \textcolor{Blue}{5},\, \textcolor{Blue}{6},\, \textcolor{Red}{7}
    \end{aligned}\\
\definecolor{Blue}{rgb}{0.29,0.37,0.89}
\definecolor{Red}{rgb}{0.82,0.01,0.11}
    \textcolor{Blue}{E(3) = \{0, 3, 5, 6\}}, \quad \textcolor{Red}{O(3) = \{1, 2, 4, 7\}}.
\end{gather}

\begin{equation}
\definecolor{Blue}{rgb}{0.29,0.37,0.89}
\definecolor{Red}{rgb}{0.82,0.01,0.11}
\begin{aligned}
    \textcolor{Blue}{0}^0 + \textcolor{Blue}{3}^0 + \textcolor{Blue}{5}^0 + \textcolor{Blue}{6}^0 = \textcolor{Red}{1}^0 + \textcolor{Red}{2}^0 +\textcolor{Red}{ 4}^0 +\textcolor{Red}{ 7}^0 &= 4,\\
    \textcolor{Blue}{0}^1 + \textcolor{Blue}{3}^1 + \textcolor{Blue}{5}^1 + \textcolor{Blue}{6}^1 = \textcolor{Red}{1}^1 + \textcolor{Red}{2}^1 + \textcolor{Red}{4}^1 + \textcolor{Red}{7}^1 &= 14,\\
    \textcolor{Blue}{0}^2 + \textcolor{Blue}{3}^2 + \textcolor{Blue}{5}^2 + \textcolor{Blue}{6}^2 = \textcolor{Red}{1}^2 + \textcolor{Red}{2}^2 + \textcolor{Red}{4}^2 + \textcolor{Red}{7}^2 &= 70.
\end{aligned}
\end{equation}
While,
\begin{equation}
\definecolor{Blue}{rgb}{0.29,0.37,0.89}
\definecolor{Red}{rgb}{0.82,0.01,0.11}
    \textcolor{Blue}{0}^3 + \textcolor{Blue}{3}^3 + \textcolor{Blue}{5}^3 + \textcolor{Blue}{6}^3 = 368 \neq \textcolor{Red}{1}^3 + \textcolor{Red}{2}^3 + \textcolor{Red}{4}^3 + \textcolor{Red}{7}^3 = 416.
\end{equation}
\end{example}

\begin{example} For $N=4$, one has
\begin{gather}
\definecolor{Blue}{rgb}{0.29,0.37,0.89}
\definecolor{Red}{rgb}{0.82,0.01,0.11}
    \begin{aligned}
    T_4 = \{&\textcolor{Blue}{0},\, \textcolor{Red}{1},\, \textcolor{Red}{1},\, \textcolor{Blue}{0},\, \textcolor{Red}{1},\, \textcolor{Blue}{0},\, \textcolor{Blue}{0},\, \textcolor{Red}{1},\, \textcolor{Red}{1},\, \textcolor{Blue}{0},\, \, \, \textcolor{Blue}{0},\, \, \, \textcolor{Red}{1},\quad \textcolor{Blue}{0},\, \, \, \textcolor{Red}{1},\, \, \, \textcolor{Red}{1},\, \, \, \textcolor{Blue}{0}\}\\
    \text{index }\, & \textcolor{Blue}{0},\, \textcolor{Red}{1},\, \textcolor{Red}{2},\, \textcolor{Blue}{3},\, \textcolor{Red}{4},\, \textcolor{Blue}{5},\, \textcolor{Blue}{6},\, \textcolor{Red}{7},\, \textcolor{Red}{8},\, \textcolor{Blue}{9},\, \textcolor{Blue}{10},\, \textcolor{Red}{11},\, \textcolor{Blue}{12},\, \textcolor{Red}{13},\, \textcolor{Red}{14},\, \textcolor{Blue}{15}
    \end{aligned}\\
\definecolor{Blue}{rgb}{0.29,0.37,0.89}
\definecolor{Red}{rgb}{0.82,0.01,0.11}
    \textcolor{Blue}{E(4) = \{0, 3, 5, 6, 9, 10, 12, 15\}}, \quad \textcolor{Red}{O(4) = \{1, 2, 4, 7, 8, 11, 13, 14\}}.
\end{gather}

\begin{equation}
\definecolor{Blue}{rgb}{0.29,0.37,0.89}
\definecolor{Red}{rgb}{0.82,0.01,0.11}
\begin{aligned}
    &\begin{multlined}
    \textcolor{Blue}{0}^0 + \textcolor{Blue}{3}^0 + \textcolor{Blue}{5}^0 + \textcolor{Blue}{6}^0 +\textcolor{Blue}{9}^0 + \textcolor{Blue}{10}^0 + \textcolor{Blue}{12}^0 + \textcolor{Blue}{15}^0 = \\
    \textcolor{Red}{1}^0 + \textcolor{Red}{2}^0 + \textcolor{Red}{4}^0 + \textcolor{Red}{7}^0 + \textcolor{Red}{8}^0 + \textcolor{Red}{11}^0 + \textcolor{Red}{13}^0 + \textcolor{Red}{14}^0 = 8
    \end{multlined}\\
    &\begin{multlined}
    \textcolor{Blue}{0}^1 + \textcolor{Blue}{3}^1 + \textcolor{Blue}{5}^1 + \textcolor{Blue}{6}^1 +\textcolor{Blue}{9}^1 + \textcolor{Blue}{10}^1 + \textcolor{Blue}{12}^1 + \textcolor{Blue}{15}^1 = \\
    \textcolor{Red}{1}^1 + \textcolor{Red}{2}^1 + \textcolor{Red}{4}^1 + \textcolor{Red}{7}^1 + \textcolor{Red}{8}^1 + \textcolor{Red}{11}^1 + \textcolor{Red}{13}^1 + \textcolor{Red}{14}^1 = 60
    \end{multlined}\\
    &\begin{multlined}
    \textcolor{Blue}{0}^2 + \textcolor{Blue}{3}^2 + \textcolor{Blue}{5}^2 + \textcolor{Blue}{6}^2 +\textcolor{Blue}{9}^2 + \textcolor{Blue}{10}^2 + \textcolor{Blue}{12}^2 + \textcolor{Blue}{15}^2 = \\
    \textcolor{Red}{1}^2 + \textcolor{Red}{2}^2 + \textcolor{Red}{4}^2 + \textcolor{Red}{7}^2 + \textcolor{Red}{8}^2 + \textcolor{Red}{11}^2 + \textcolor{Red}{13}^2 + \textcolor{Red}{14}^2 = 620
    \end{multlined}\\
    &\begin{multlined}
    \textcolor{Blue}{0}^3 + \textcolor{Blue}{3}^3 + \textcolor{Blue}{5}^3 + \textcolor{Blue}{6}^3 +\textcolor{Blue}{9}^3 + \textcolor{Blue}{10}^3 + \textcolor{Blue}{12}^3 + \textcolor{Blue}{15}^3 = \\
    \textcolor{Red}{1}^3 + \textcolor{Red}{2}^3 + \textcolor{Red}{4}^3 + \textcolor{Red}{7}^3 + \textcolor{Red}{8}^3 + \textcolor{Red}{11}^3 + \textcolor{Red}{13}^3 + \textcolor{Red}{14}^3 = 7200
    \end{multlined}
\end{aligned}
\end{equation}

\end{example}

This relation demonstrates that the Thue-Morse sequence provides a solution to the Prouhet-Tarry-Escott problem (or multigrades problem) for the given $k$ \cite{TMmaths}.

\subsection{Definition of PTM (logical) states}
First, on top of the notations defined in \ref{apdx:def} we define the notation $\ket{(k)_2}$, representing the $N$ qubits state indexed by the base $2$ notation of $k$ with $N$ digits, such that, for example, $\ket{(0)_2}=\ket*{\underbrace{00\dots0}_{N \text{ times}}}$ or $\ket{(2^N-1)_2}=\ket*{\underbrace{11\dots1}_{N \text{ times}}}$ \todo*{$(k)_2\equiv ?$}. 

\begin{definition}
We define the \textit{PTM (logical) states} on a $N$ qubits system, in the uncoupled basis, as: 

\begin{equation}
     \ket{1_\text{TM}^{(N)}} = \frac{1}{\sqrt{2^{N-1}}} \sum_{k=0}^{2^N-1} t_k \ket{(k)_2} = \frac{1}{\sqrt{2^{N-1}}} \sum_{o\in O(N)}\ket{(o)_2},
\end{equation}

and its complementary (using again $\overline{t} \equiv 1-t$)\todo*{redefine $\overline{t}$} :

\begin{equation}
     \ket{0_\text{TM}^{(N)}} = \frac{1}{\sqrt{2^{N-1}}}\sum_{k=0}^{2^N-1} \bar{t_k} \ket{(k)_2}= \frac{1}{\sqrt{2^{N-1}}} \sum_{e\in E(N)}\ket{(e)_2}.
\end{equation}

\end{definition}

Note the ${1}/{\sqrt{2^{N-1}}}$ prefactor, it arises simply because, by construction, in the first $2^N$ elements of the PTM sequence, exactly half are $1$'s \todo*{Explain prefactor N-1}.

If one notes that the PTM sequence $t_i$ at index $i$ is equal to $1$ if and only if the binary representation of $i$ has an odd amount of $1$'s, we get trivially the following property:
$\forall N \in \mathbb{N^*},$ $$\ket{0_\text{TM}^{(N+1)}} = \frac{1}{\sqrt{2}}\left(\ket{0}\otimes\ket{0_\text{TM}^{(N)}} + \ket{1}\otimes \ket{1_\text{TM}^{(N)}}\right)$$and$$\ket{1_\text{TM}^{(N+1)}} = \frac{1}{\sqrt{2}}\left(\ket{0}\otimes\ket{1_\text{TM}^{(N)}} + \ket{1}\otimes \ket{0_\text{TM}^{(N)}}\right)$$ 

Here is an example of construction, the states contributing to $\ket{0_\text{TM}^{(2)}}$ (blue) and $\ket{1_\text{TM}^{(2)}}$ (red) are highlighted.

\begin{tikzpicture}[x=0.75pt,y=0.75pt,yscale=-1,xscale=1]

\draw (60,67.78) node [anchor=north west][inner sep=0.75pt]  [color={rgb, 255:red, 74; green, 94; blue, 226 }  ,opacity=1 ]  {$E( 2) \ =\ \{0\ \ \ \ \ \ \ \ \ \ \ \ 3\}$};
\draw (41,92.78) node [anchor=north west][inner sep=0.75pt]  [color={rgb, 255:red, 74; green, 94; blue, 226 }  ,opacity=1 ]  {$( E( 2))_{2} \ =\ \{00\ \ \ \ \ \ \ \ \ 11\}$};
\draw (60,151.65) node [anchor=north west][inner sep=0.75pt]  [color={rgb, 255:red, 208; green, 2; blue, 27 }  ,opacity=1 ]  {$O( 2) \ =\ \{\ \ \ \ 1\ \ \ \ 2\ \ \ \ \}$};
\draw (41,176.65) node [anchor=north west][inner sep=0.75pt]  [color={rgb, 255:red, 208; green, 2; blue, 27 }  ,opacity=1 ]  {$( O( 2))_{2} \ =\ \{\ \ \ 01\ \ \ 10\ \ \ \}$};
\draw (230,149.15) node [anchor=north west][inner sep=0.75pt]  [color={rgb, 255:red, 208; green, 2; blue, 27 }  ,opacity=1 ]  {$\Longrightarrow \ket{1_\text{TM}^{( 2)}} =\frac{1}{\sqrt{2}}\left(\ket{01} +\ket{10}\right)$};
\draw (77,11.4) node [anchor=north west][inner sep=0.75pt]    {$T_{2} \ =\ \{\textcolor[rgb]{0.29,0.37,0.89}{0} \ \ \ \textcolor[rgb]{0.82,0.01,0.11}{1\ \ \ 1} \ \ \ \textcolor[rgb]{0.29,0.37,0.89}{0}\}$};
\draw (79,36.4) node [anchor=north west][inner sep=0.75pt]    {$\textcolor[rgb]{0.6,0.6,0.6}{\ \ \ \ \ \ \ \ \ \ 0 \ \ \ 1 \ \ \ 2 \ \ \ 3}$};
\draw (230,65.28) node [anchor=north west][inner sep=0.75pt]  [color={rgb, 255:red, 74; green, 94; blue, 226 }  ,opacity=1 ]  {$\Longrightarrow \ket{0_\text{TM}^{( 2)}} =\frac{1}{\sqrt{2}}\left(\ket{00} +\ket{11}\right)$};
\draw (1,4.4) node [anchor=north west][inner sep=0.75pt]    {$\boxed{N=2}$};

\end{tikzpicture}   

Moreover, by adding one more qubit, one can iterate the construction of, this time, $\ket{0_\text{TM}^{(3)}}$ and $\ket{1_\text{TM}^{(3)}}$. The components of the $\ket{0_\text{TM}^{(2)}}$ (resp. $\ket{1_\text{TM}^{(2)}}$) PTM state of the two qubit subsystem are underlined in blue (resp. red).

\begin{tikzpicture}[x=0.75pt,y=0.75pt,yscale=-1,xscale=1]

\definecolor{Blue}{rgb}{0.29,0.37,0.89}
\definecolor{Red}{rgb}{0.82,0.01,0.11}

\draw (75,12.4) node [anchor=north west][inner sep=0.75pt]    {$T_{3} \ =\ \{0\ \ \ 1\ \ \ 1\ \ \ 0\ \ \ 1\ \ \ 0\ \ \ 0\ \ \ 1\}$};
\draw (77,36.4) node [anchor=north west][inner sep=0.75pt]    {$\textcolor[rgb]{0.6,0.6,0.6}{\ \ \ \ \ \ \ \ \ \ 0 \ \ \ 1 \ \ \ 2 \ \ \ 3 \ \ \ 4 \ \ \ 5 \ \ \ 6 \ \ \ 7}$};
\draw (64,116.28) node [anchor=north west][inner sep=0.75pt]    {$\Longrightarrow \textcolor[rgb]{0.29,0.37,0.89}{\ket{0_\text{TM}^{( 3)}} =\frac{1}{2}\left(\ket{0\text{\setulcolor{Blue}\ul{$00$}}} +\ket{0\text{\setulcolor{Blue}\ul{$11$}}} +\ket{1\text{\setulcolor{Red}\ul{$01$}}} +\ket{1\text{\setulcolor{Red}\ul{$10$}}}\right)} =\frac{1}{\sqrt{2}}\left(\ket{0} \otimes \textcolor[rgb]{0.29,0.37,0.89}{\ket{0_\text{TM}^{( 2)}}} +\ket{1} \otimes \textcolor[rgb]{0.82,0.01,0.11}{\ket{1_\text{TM}^{( 2)}}}\right)$};
\draw (64,223.65) node [anchor=north west][inner sep=0.75pt]    {$\Longrightarrow \textcolor[rgb]{0.82,0.01,0.11}{\ket{1_\text{TM}^{(3)}} =\frac{1}{2}\left(\ket{0\text{\setulcolor{Red}\ul{$01$}}} +\ket{0\text{\setulcolor{Red}\ul{$10$}}} +\ket{1\text{\setulcolor{Blue}\ul{$00$}}} +\ket{1\text{\setulcolor{Blue}\ul{$11$}}}\right)} =\frac{1}{\sqrt{2}}\left(\ket{0} \otimes \textcolor[rgb]{0.82,0.01,0.11}{\ket{1_\text{TM}^{( 2)}}} +\ket{1} \otimes \textcolor[rgb]{0.29,0.37,0.89}{\ket{0_\text{TM}^{( 2)}}}\right)$};
\draw (1,4.4) node [anchor=north west][inner sep=0.75pt]    {$\boxed{N=3}$};
\draw (44,93.78) node [anchor=north west][inner sep=0.75pt]    {\textcolor[rgb]{0.29,0.37,0.89}{$( E( 3))_{2} \ =\ \{0$\setulcolor{Blue}\ul{$00$}$ \ \ \ \ \ \ \ 0$\setulcolor{Blue}\ul{$11$}$ \ \ \ \ 1$\setulcolor{Red}\ul{$01$}$\ 1$\setulcolor{Red}\ul{$10$}$\ \ \}$}};
\draw (63,68.78) node [anchor=north west][inner sep=0.75pt]    {$E( 3) \ =\ \{0\ \ \ \ \ \ \ \ \ \ \ 3\ \ \ \ \ \ \ \ 5\ \ \ 6\ \ \ \ \}$};
\draw (42,200.65) node [anchor=north west][inner sep=0.75pt]    {\textcolor[rgb]{0.82,0.01,0.11}{$( O( 3))_{2} \ =\ \{\ \ 0$\setulcolor{Red}\ul{$01$}$\  0$\setulcolor{Red}\ul{$10$}$ \ \ \ \ 1$\setulcolor{Blue}\ul{$00$}$ \ \ \ \ \ \ \ 1$\setulcolor{Blue}\ul{$11$}$\}$}};
\draw (60,175.65) node [anchor=north west][inner sep=0.75pt]    {$O( 3) \ =\ \{\ \ \ \ 1\ \ \ 2\ \ \ \ \ \ \ \ 4\ \ \ \ \ \ \ \ \ \ \ 7\}$};

\end{tikzpicture}

\begin{definition}
    We define the qubit spin (Pauli) operators: 
    \begin{equation}
    \sigma_x = \begin{pmatrix} 0 & 1\\ 1 & 0 \end{pmatrix}, \quad \sigma_y = \begin{pmatrix} 0 & -i\\ i & 0 \end{pmatrix}, \quad \sigma_z = \begin{pmatrix} 1 & 0\\ 0 & -1 \end{pmatrix}.
\end{equation}

We will use $\sigma_i^{(k)}$ to indicate that the operator $\sigma_i$ acts on the $k$-th qubit among $N$ qubits, i.e. $\sigma_i^{(k)} = \underbrace{\mathbb{1}_2 \otimes \cdots  \otimes \mathbb{1}_2}_{k-1\text{ times}} \otimes \, \sigma_i \otimes \underbrace{\mathbb{1}_2 \otimes \cdots \otimes \mathbb{1}_2}_{N-k \text{ times}}$.
\end{definition}

The PTM states present a certain number of interesting properties regarding spin operators $\sigma_{x,y,z}$:

\begin{enumerate}[label=\textbf{Property \thesubsection.\arabic*},leftmargin=*]
    \item $\forall N>1,  \ev**{\sum_{k=1}^N \sigma_{x,y,z}^{(k)}}{0_\text{TM}^{(N)}} = 0$ and $\ev**{\sum_{k=1}^N \sigma_{x,y,z}^{(k)}}{1_\text{TM}^{(N)}} = 0$.
    \item Let $Z$ be an ensemble of $\sigma_z^{(k)}$ acting on $M$ qubits ($M<N$) \\ $$\ev**{\prod_{k\in Z} \sigma_z^{(k)}}{0_\text{TM}^{(N)}} = \ev**{\prod_{k\in Z} \sigma_z^{(k)}}{1_\text{TM}^{(N)}},$$ $$\text{and }\mel**{1_\text{TM}^{(N)}}{\prod_{k\in Z} \sigma_z^{(k)}}{0_\text{TM}^{(N)}} = 0.$$
    \item Moreover $\forall j < N, i\in\{y,z\}$ \todo*{check for $\sigma_y$} $$\ev**{\left(\sum_{k=1}^N \sigma_{i}^{(k)}\right)^{j}}{0_\text{TM}^{(N)}} = \ev**{\left(\sum_{k=1}^N \sigma_{i}^{(k)}\right)^{j}}{1_\text{TM}^{(N)}},$$ $$\text{and }\mel**{1_\text{TM}^{(N)}}{\left(\sum_{k=1}^N \sigma_{i}^{(k)}\right)^{j}}{0_\text{TM}^{(N)}} = 0.$$
    \item $\forall k,j < N,$ $\quad \sigma_x^{(k)}\sigma_x^{(j)}\ket{0_\text{TM}^{(N)}} = \ket{0_\text{TM}^{(N)}} $ and $\sigma_x^{(k)}\sigma_x^{(j)}\ket{1_\text{TM}^{(N)}} = \ket{1_\text{TM}^{(N)}} $.
\end{enumerate}

We will see in the following sections how these properties can be used for various applications.

\subsection{Relations between the PTM states and the Hadamard gate}

\begin{definition}
    We define the single qubit Hadamard gate, acting on the $k$-th qubit among $N$ qubits:
\begin{equation}
    H^{(k)} = \underbrace{\mathbb{1}_2 \otimes \cdots \otimes \mathbb{1}_2}_{k-1\text{ times}} \otimes \frac{1}{\sqrt{2}}\begin{pmatrix}
        1 & 1 \\ 1 & -1
    \end{pmatrix} \otimes \underbrace{\mathbb{1}_2 \otimes \cdots \otimes \mathbb{1}_2}_{N-k \text{ times}}
\end{equation}
\end{definition}

\begin{property}
The first interesting property of the Hadamard gate relating to these newly defined PTM states is that 
\begin{equation}
\label{eq::HTM}
    \begin{aligned}
        \bigotimes_{k=1}^N H^{(k)} \ket{(0)_2} &= \frac{1}{\sqrt{2}} \left(\ket{0_\text{TM}^{(N)}} + \ket{1_\text{TM}^{(N)}} \right)\\
        \bigotimes_{k=1}^N H^{(k)} \ket{(2^N-1)_2} &= \frac{1}{\sqrt{2}} \left(\ket{0_\text{TM}^{(N)}} - \ket{1_\text{TM}^{(N)}} \right),
    \end{aligned}
\end{equation}
allowing for a simple and natural passage from the PTM states to the uncoupled basis, namely

\begin{equation}
    \begin{aligned}
        \bigotimes_{k=1}^N H^{(k)} \ket{0_\text{TM}^{(N)}} &= \frac{1}{\sqrt{2}} \left(\ket{(0)_2} + \ket{(2^N-1)_2} \right)\\
        \bigotimes_{k=1}^N H^{(k)} \ket{1_\text{TM}^{(N)}} &= \frac{1}{\sqrt{2}} \left(\ket{(0)_2} - \ket{(2^N-1)_2} \right).
    \end{aligned}
\end{equation} \todo*{show realization example avec circuit quantique par exemple}
\end{property}

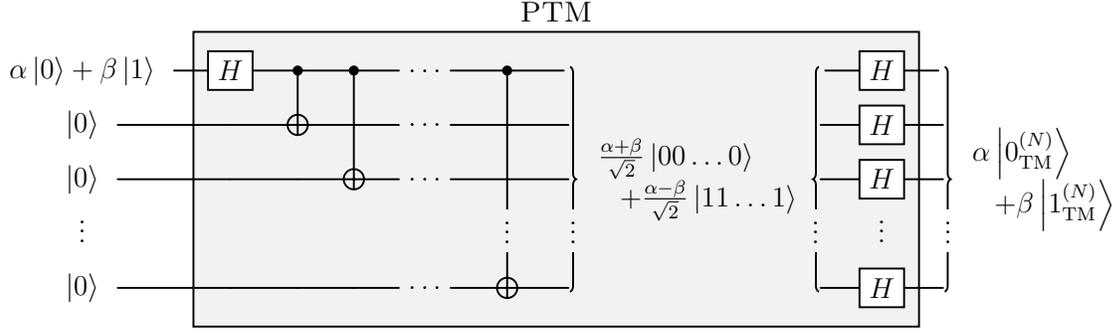
\begin{figure}[hbtp!]
    \centering
    \begin{minipage}{\textwidth}
        \resizebox{\textwidth}{!}{
\begin{quantikz}[ row sep={0.8cm,between origins}, wire types={q,q,q,n,q}]
\midstick[wires=1,brackets=right]{$\alpha \ket{0} + \beta \ket{1}$} & \gate{H} \gategroup[5,steps=7,style={inner
xsep=2pt, fill=gray!10},background, label style={label
position=above, anchor=south, yshift=-.2cm}]{\sc PTM} & \ctrl{1} & \ctrl{2} & \ \ldots\ & \ctrl{4} & \midstick[wires=5]{$\begin{matrix}\ \frac{\alpha+\beta}{\sqrt{2}}\ket{00\dots0}\ \quad \ \\ \quad\ +\frac{\alpha-\beta}{\sqrt{2}}\ket{11\dots1}\ \end{matrix}$} & \gate{H} & \midstick[wires=5, brackets=left]{$\begin{matrix}\ \alpha\ket{0_\text{TM}^{(N)}}\ \quad \ \\ \quad\ +\beta\ket{1_\text{TM}^{(N)}}\ \end{matrix}$}\\
\midstick[wires=1,brackets=right]{$\ket{0}$} & & \targ{} & & \ \ldots\ &  & & \gate{H} & \\
\midstick[wires=1,brackets=right]{$\ket{0}$} & & & \targ{} &  \ \ldots\ & & & \gate{H} & \\
\midstick[wires=1,brackets=none, label style={yshift=4pt}]{$\vdots$} & & & & & \midstick[wires=1,brackets=none, label style={yshift=1pt, fill=gray!10, text height=10pt}]{$\vdots$} & \lstick[wires=1,brackets=none, label style={yshift=1pt, xshift=8pt, fill=gray!10, text height=10pt}]{$\vdots$}\rstick[wires=1,brackets=none, label style={yshift=1pt, xshift=-8pt, fill=gray!10, text height=10pt}]{$\vdots$}& \midstick[wires=1,brackets=none, label style={yshift=4pt}]{$\vdots$} & \lstick[wires=1,brackets=none, label style={yshift=1pt, xshift=8pt, fill=white, text height=10pt}]{$\vdots$}\\
\midstick[wires=1,brackets=right]{$\ket{0}$} & & & & \ \ldots\ & \targ{} & & \gate{H} &
\end{quantikz}}
    \end{minipage}
    \caption{Definition of the PTM Gate acting on $N$ qubits.}
    \label{fig:TMgate}
\end{figure}

An example of a circuit for encoding a state $\ket{\psi} = \alpha \ket{0} + \beta \ket{1}$ into $\ket{\psi_\text{TM}} = \alpha \ket{0_\text{TM}^{(N)}} + \beta \ket{1_\text{TM}^{(N)}}$ is given in Fig.\ref{fig:TMgate}.

\subsection{$X$-$X$ Spin chain and the PTM states}

Property 1.2.4 can be generalized to verify that the PTM states are among the eigenstates of $X$-$X$ rotations, and share the same eigenvalue:
$$\forall k,j < N,\quad e^{i\theta\sigma_x^{(k)}\sigma_x^{(j)}}\ket{0_\text{TM}^{(N)}} = e^{i\theta}\ket{0_\text{TM}^{(N)}} \text{ and } e^{i\theta\sigma_x^{(k)}\sigma_x^{(j)}}\ket{1_\text{TM}^{(N)}} = e^{i\theta}\ket{1_\text{TM}^{(N)}} $$ \todo*{séparer en deux équations}
This means that $X$-$X$ interactions only add a global phase in the Thue-Morse basis, making the encoding particularly useful to store memory qubits in $X$-$X$ PTM spin lattices or chains.

\subsection{PTM states as eigenvalues of $S_x$ \label{ssec:SXTM}}

By defining $S_{x,z} = \sum_{k=1}^N \sigma_{x,z}^{(k)}$, we have that
\begin{equation}
    S_z\ket{(0)_2} = -N\ket{(0)_2} \;\text{ and }\; S_z\ket{(2^N-1)_2} = N\ket{(2^N-1)_2}.
\end{equation}

Using (\ref{eq::HTM}), the relation $H^{(k)}\sigma_z^{(k)}H^{(k)}=\sigma_x^{(k)}$ and consenquently $\left(\bigotimes_{k=1}^N H^{(k)}\right) S_z \left(\bigotimes_{k=1}^N H^{(k)}\right) = S_x$, it is easy to see that
\begin{equation}
    S_x\ket{0_\text{TM}^{(N)}} = N\ket{1_\text{TM}^{(N)}} \;\text{ and }\; S_x\ket{1_\text{TM}^{(N)}} = N\ket{0_\text{TM}^{(N)}}.
\end{equation}

The states $\frac{1}{\sqrt{2}}\left(\ket{0_\text{TM}^{(N)}}\pm \ket{1_\text{TM}^{(N)}}\right)$ are therefore eigenstates of $S_x$ with eigenvalues $\pm N$. 

\section{Appearance of the PTM sequence in the Hilbert spaces of given quantum computing platforms}

\subsection{The PTM sequence as the indicator function of the purely dephased $X$-$X$ Ising chain}

\begin{definition}
We define an $X$-$X$ Ising chain of decohering qubits as a system of $N$ qubits evolving according to the equation:

\begin{equation}\label{eq:XX}
\begin{aligned}
    \forall k \in [1,N],  &\quad L_k = \underbrace{\mathbb{1}_2^{(1)}\otimes\dots\otimes\mathbb{1}_2^{(k-1)}}_{k-1} \otimes \; S_z^{(k)} \otimes \underbrace{\mathbb{1}_2^{(k+1)}\otimes\dots\otimes\mathbb{1}_2^{(N)}}_{N-k}, \\
    \partial_t \rho (t) &= -\frac{i}{\hbar} \left[H(t),\rho(t)\right] + \sum_{k=1}^N\gamma_k\left( L_k \rho(t) L_k^\dag - \frac{1}{2} \left\{ L_k^\dag L_k, \rho(t)\right\}\right).
\end{aligned}
\end{equation}

{With $H(t) = g \sum_{k=1}^{N-1} S^{(k)}_x \otimes S^{(k+1)}_x$ \todo*{more than just chain I think it also works in a network? NO}, and $\rho(t)$ the density matrix representing the state of the chain at time $t$.} Here, \(L_k\) represents the Lindblad operator corresponding to the decoherence of the \(k\)-th qubit in the chain, with \(S_z^{(k)} = \frac{1}{2} \sigma_z^{(k)}\), and \(\gamma_k\) is the decoherence rate associated with the \(k\)-th qubit.
\end{definition}

If the system is initialized in the pure eigenstate $\rho(0) = \ket{i_0}\bra{i_0}$, where $i_0$ belongs to either the set $E(N)$ or $O(N)$, then the final state will be a statistical mixture of states within the same set as the initial state. This implies that the PTM sequence acts as an indicator function, determining whether a state is present in the statistical mixture at the end of the evolution. In other words, $\forall t', \; \rho_{i,k}(t') \neq 0$ if $t_i = t_k = t_{i_0}$.

The linearity of time evolution allows us to extend this concept: if the initial state is a linear combination of eigenstates, all of which are elements of $E(N)$ (or $O(N)$), then the final state will also be a statistical mixture of states in $E(N)$ (or $O(N)$). Furthermore, if the initial state includes eigenstates from both sets, the final state will be a mixture of all eigenstates.

\begin{figure}[hbtp!]
    \centering
    \fcolorbox{black}{white}{
    {
  
\tikzset {_wvbdw6u59/.code = {\pgfsetadditionalshadetransform{ \pgftransformshift{\pgfpoint{89.1 bp } { -108.9 bp }  }  \pgftransformscale{1.32 }  }}}
\pgfdeclareradialshading{_2n2lra8le}{\pgfpoint{-72bp}{88bp}}{rgb(0bp)=(1,1,1);
rgb(0bp)=(1,1,1);
rgb(25bp)=(0,0,0);
rgb(400bp)=(0,0,0)}

  
\tikzset {_tgfkp3rmg/.code = {\pgfsetadditionalshadetransform{ \pgftransformshift{\pgfpoint{89.1 bp } { -108.9 bp }  }  \pgftransformscale{1.32 }  }}}
\pgfdeclareradialshading{_71mcwwg3o}{\pgfpoint{-72bp}{88bp}}{rgb(0bp)=(1,1,1);
rgb(0bp)=(1,1,1);
rgb(25bp)=(0,0,0);
rgb(400bp)=(0,0,0)}

  
\tikzset {_zd57t6wj0/.code = {\pgfsetadditionalshadetransform{ \pgftransformshift{\pgfpoint{89.1 bp } { -108.9 bp }  }  \pgftransformscale{1.32 }  }}}
\pgfdeclareradialshading{_04whbmxca}{\pgfpoint{-72bp}{88bp}}{rgb(0bp)=(1,1,1);
rgb(0bp)=(1,1,1);
rgb(25bp)=(0,0,0);
rgb(400bp)=(0,0,0)}

  
\tikzset {_7azmdj2aw/.code = {\pgfsetadditionalshadetransform{ \pgftransformshift{\pgfpoint{89.1 bp } { -108.9 bp }  }  \pgftransformscale{1.32 }  }}}
\pgfdeclareradialshading{_we3a9wvpi}{\pgfpoint{-72bp}{88bp}}{rgb(0bp)=(1,1,1);
rgb(0bp)=(1,1,1);
rgb(25bp)=(0,0,0);
rgb(400bp)=(0,0,0)}

  
\tikzset {_xt6mtlr6c/.code = {\pgfsetadditionalshadetransform{ \pgftransformshift{\pgfpoint{89.1 bp } { -108.9 bp }  }  \pgftransformscale{1.32 }  }}}
\pgfdeclareradialshading{_9engvswtg}{\pgfpoint{-72bp}{88bp}}{rgb(0bp)=(1,1,1);
rgb(0bp)=(1,1,1);
rgb(25bp)=(0,0,0);
rgb(400bp)=(0,0,0)}

  
\tikzset {_9mjcxqe6q/.code = {\pgfsetadditionalshadetransform{ \pgftransformshift{\pgfpoint{89.1 bp } { -108.9 bp }  }  \pgftransformscale{1.32 }  }}}
\pgfdeclareradialshading{_6vio8wz5f}{\pgfpoint{-72bp}{88bp}}{rgb(0bp)=(1,1,1);
rgb(0bp)=(1,1,1);
rgb(25bp)=(0,0,0);
rgb(400bp)=(0,0,0)}
\tikzset{every picture/.style={line width=0.75pt}} 

\begin{tikzpicture}[x=0.75pt,y=0.75pt,yscale=-1,xscale=1]
\useasboundingbox (85, 0) rectangle (\textwidth, 180);

\draw [color={rgb, 255:red, 255; green, 0; blue, 0 }  ,draw opacity=1 ][line width=2.25]    (333.1,32.41) -- (358.63,32.41) (344.6,28.41) -- (344.6,36.41)(356.1,28.41) -- (356.1,36.41) ;
\draw [color={rgb, 255:red, 255; green, 0; blue, 0 }  ,draw opacity=1 ][line width=2.25]    (212.09,32.41) -- (297,32.41) (223.59,28.41) -- (223.59,36.41)(235.09,28.41) -- (235.09,36.41)(246.59,28.41) -- (246.59,36.41)(258.09,28.41) -- (258.09,36.41)(269.59,28.41) -- (269.59,36.41)(281.09,28.41) -- (281.09,36.41)(292.59,28.41) -- (292.59,36.41) ;
\draw [color={rgb, 255:red, 255; green, 0; blue, 0 }  ,draw opacity=1 ][line width=2.25]    (317.25,32.41) -- (329.25,32.41) (321.75,28.41) -- (321.75,36.41) ;

\draw   (238,107.24) -- (326.18,107.24) -- (326.18,96.49) -- (351.5,118) -- (326.18,139.51) -- (326.18,128.76) -- (238,128.76) -- cycle ;

\draw [line width=1.5]    (373,83.01) .. controls (374.67,84.68) and (374.67,86.34) .. (373,88.01) .. controls (371.33,89.68) and (371.33,91.34) .. (373,93.01) .. controls (374.67,94.68) and (374.67,96.34) .. (373,98.01) .. controls (371.33,99.68) and (371.33,101.34) .. (373,103.01) .. controls (374.67,104.68) and (374.67,106.34) .. (373,108.01) .. controls (371.33,109.68) and (371.33,111.34) .. (373,113.01) .. controls (374.67,114.68) and (374.67,116.34) .. (373,118.01) .. controls (371.33,119.68) and (371.33,121.34) .. (373,123.01) .. controls (374.67,124.68) and (374.67,126.34) .. (373,128.01) .. controls (371.33,129.68) and (371.33,131.34) .. (373,133.01) .. controls (374.67,134.68) and (374.67,136.34) .. (373,138.01) .. controls (371.33,139.68) and (371.33,141.34) .. (373,143.01) .. controls (374.67,144.68) and (374.67,146.34) .. (373,148.01) -- (373,152.99) -- (373,152.99) ;

\draw  [draw opacity=0][shading=_2n2lra8le,_wvbdw6u59] (204.18,32.41) .. controls (204.18,28.04) and (207.72,24.5) .. (212.09,24.5) .. controls (216.46,24.5) and (220,28.04) .. (220,32.41) .. controls (220,36.78) and (216.46,40.32) .. (212.09,40.32) .. controls (207.72,40.32) and (204.18,36.78) .. (204.18,32.41) -- cycle ;
\draw  [draw opacity=0][shading=_71mcwwg3o,_tgfkp3rmg] (227.18,32.41) .. controls (227.18,28.04) and (230.72,24.5) .. (235.09,24.5) .. controls (239.46,24.5) and (243,28.04) .. (243,32.41) .. controls (243,36.78) and (239.46,40.32) .. (235.09,40.32) .. controls (230.72,40.32) and (227.18,36.78) .. (227.18,32.41) -- cycle ;
\draw  [draw opacity=0][shading=_04whbmxca,_zd57t6wj0] (250.18,32.41) .. controls (250.18,28.04) and (253.73,24.5) .. (258.1,24.5) .. controls (262.47,24.5) and (266.01,28.04) .. (266.01,32.41) .. controls (266.01,36.78) and (262.47,40.32) .. (258.1,40.32) .. controls (253.73,40.32) and (250.18,36.78) .. (250.18,32.41) -- cycle ;
\draw  [draw opacity=0][shading=_we3a9wvpi,_7azmdj2aw] (273.18,32.41) .. controls (273.18,28.04) and (276.72,24.5) .. (281.09,24.5) .. controls (285.46,24.5) and (289,28.04) .. (289,32.41) .. controls (289,36.78) and (285.46,40.32) .. (281.09,40.32) .. controls (276.72,40.32) and (273.18,36.78) .. (273.18,32.41) -- cycle ;
\draw  [draw opacity=0][shading=_9engvswtg,_xt6mtlr6c] (325.18,32.41) .. controls (325.18,28.04) and (328.73,24.5) .. (333.1,24.5) .. controls (337.47,24.5) and (341.01,28.04) .. (341.01,32.41) .. controls (341.01,36.78) and (337.47,40.32) .. (333.1,40.32) .. controls (328.73,40.32) and (325.18,36.78) .. (325.18,32.41) -- cycle ;
\draw  [draw opacity=0][shading=_6vio8wz5f,_9mjcxqe6q] (348.18,32.41) .. controls (348.18,28.04) and (351.72,24.5) .. (356.09,24.5) .. controls (360.46,24.5) and (364,28.04) .. (364,32.41) .. controls (364,36.78) and (360.46,40.32) .. (356.09,40.32) .. controls (351.72,40.32) and (348.18,36.78) .. (348.18,32.41) -- cycle ;

\draw (296,29) node [anchor=north west][inner sep=0.75pt]   [align=left] {$\cdots$};
\draw (201,38) node [anchor=north west][inner sep=0.75pt]    {$\underbrace{\ \ \ \ \ \ \ \ \ \ \ \ \ \ \ \ \ \ \ \ \ \ \ \ \ \ \ \ \ \ \ \ }_\text{\normalsize $N$ qubits}$};
\draw (23,105.4) node [anchor=north west][inner sep=0.75pt]    {$\frac{1}{\sqrt{2}}\left(\ket{0000...00} +\ket{1111...11}\right) \ $};
\draw (24,83) node [anchor=north west][inner sep=0.75pt]   [align=left] {Initial state};
\draw (242,109.49) node [anchor=north west][inner sep=0.75pt]   [align=left] {time evolution};
\draw (327,156) node [anchor=north west][inner sep=0.75pt]   [align=left] {measurement};
\draw (435,109.4) node [anchor=north west][inner sep=0.75pt]    {$k \in O( N) \Leftrightarrow N\ \text{is odd}$};
\draw (384,83) node [anchor=north west][inner sep=0.75pt]   [align=left] {Final state $\ket{(k)_2}$};

\draw (48.25,6.31) node [anchor=north west][inner sep=0.55pt]    {$X\text{-}X\ \text{interaction}$};
\draw [color={rgb, 255:red, 255; green, 0; blue, 0 }  ,draw opacity=1 ][line width=2.25]    (18.84,14.91) -- (41.38,14.91) (30.34,10.91) -- (30.34,18.91) ;

\end{tikzpicture}
}}
    \caption{Schematic representation of the interplay between the decoherence behavior of a $X$-$X$ Ising chain initialized in the GHZ state, and the PTM sequence.}
    \label{fig::PTMevol}
\end{figure}
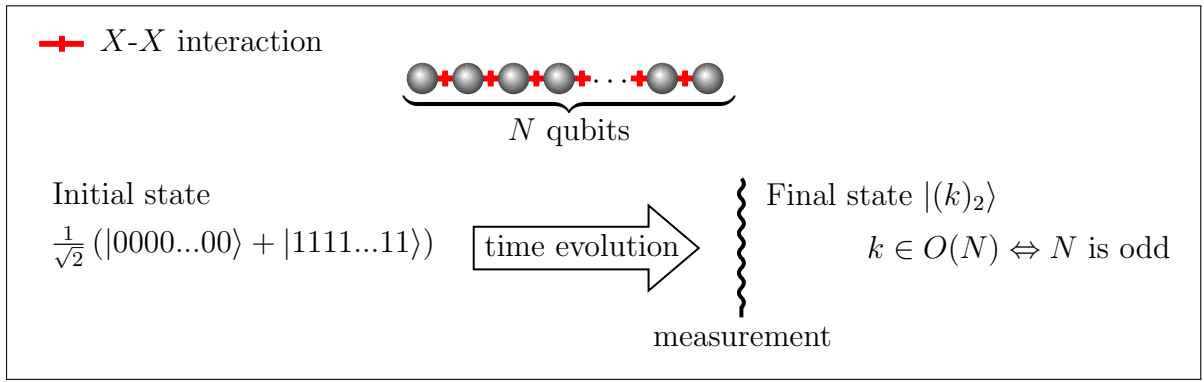

A special consideration applies to initial states of the form $\ket{\text{GHZ}} = \frac{1}{\sqrt{2}} \left(\ket{(0)_2} + \ket{(2^N - 1)_2}\right)$. By definition, $0 \in E(N)$, but $2^N - 1 \in E(N)$ if and only if $N$ is even; otherwise, $2^N-1 \in O(N)$. Therefore, an odd number of qubits initialized in a GHZ state will decohere into a statistical mixture of all eigenstates, whereas an even number of qubits will decohere into a statistical mixture of eigenstates only in $E$. See Fig. \ref{fig::PTMevol}.\todo*{add graphs}

\subsection{Initializing a spin system in a Thue-Morse state}

As we have seen in subsection \ref{ssec:SXTM}, the Thue-Morse states are mutually reciprocal under $S_x$, playing a role similar to  $\frac{1}{\sqrt{2}} \left(\ket{(0)_2} \pm \ket{(2^N-1)_2} \right)$ for $S_z$, as hinted at by (\ref{eq::HTM}). 

By using the property that, for a single qubit,
$\frac{\pi}{2T\sqrt{2}}\left(\sigma_z+\sigma_x\right)$ is the Hamiltonian that optimally generates the Hadamard gate in a time $T$ up to a global phase \cite{Aifer2022}, we get that $H^{(k)} \equiv \text{exp}\left(\frac{-i\pi}{2\sqrt{2}}\left(\sigma_z^{(k)}+\sigma_x^{(k)}\right)\right)$.

It is therefore clear that implementing the Hamiltonian $\frac{\pi}{2\sqrt{2}}\left(S_z+S_x\right)$ on a chain of spins in the $\frac{1}{\sqrt{2}} \left(\ket{(0)_2} \pm \ket{(2^N-1)_2} \right)$ state would allow to initialize a state in either $\ket{0_\text{TM}^{(N)}}$ or $\ket{1_\text{TM}^{(N)}}$, since, from (\ref{eq::HTM}) we have that

\begin{equation}
    \begin{aligned}
        \frac{1}{\sqrt{2}} \bigotimes_{k=1}^N H^{(k)}  \left(\ket{(0)_2} \pm \ket{(2^N-1)_2} \right) = \ket{0,1_\text{TM}^{(N)}}.
    \end{aligned}
\end{equation}

\begin{figure}[htbp!]
    \centering
    \includegraphics[width=\textwidth]{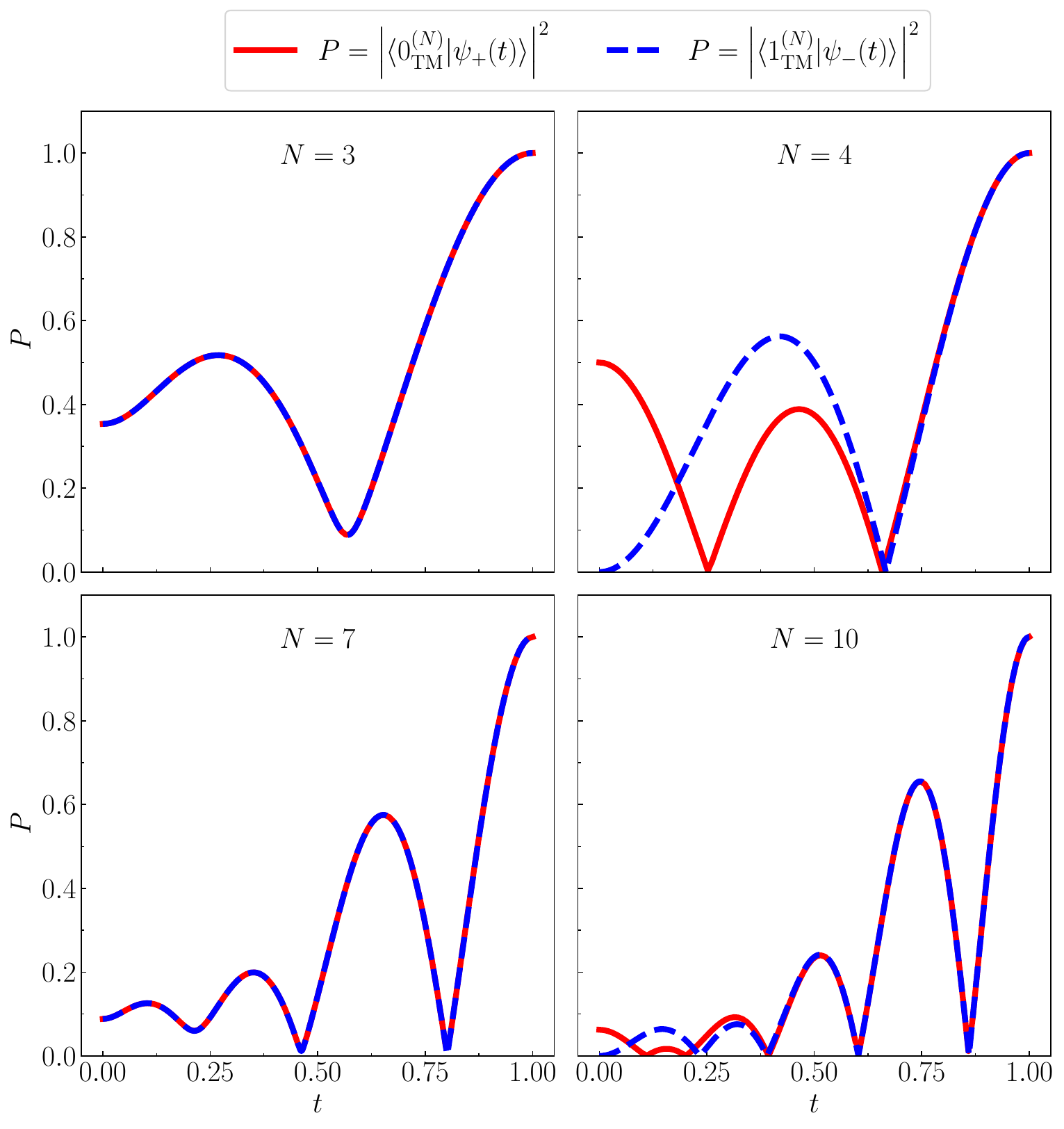}
    \caption{Time-evolution of the populations $P$ of different states initialized in $\ket{\psi_+(0)} = \frac{1}{\sqrt{2}}\left(\ket{(0)_2} + \ket{(N)_2}\right)$ (red) or $\ket{\psi_-(0)} = \frac{1}{\sqrt{2}}\left(\ket{(0)_2} - \ket{(N)_2}\right)$ (blue). With $\ket{\psi_\pm(t)} = \text{exp}\left(\frac{-i\pi}{2\sqrt{2}}\left(S_z+S_x\right)t\right)\ket{\psi_\pm(0)}$. The final state corresponds to $\ket{\psi_\pm(1)} = \bigotimes_{k=1}^N H^{(k)}\ket{\psi_\pm(0)}$.}
    \label{fig::TMevol}
\end{figure}

Examples of implementations for different $N$ are shown on Fig.\ref{fig::TMevol}.

\section{Uses of the PTM sequence in quantum computing}

\subsection{Quantum Error Correction (QEC)}
\todo*{example N=3}
Property 1.2.2 shows that the PTM states satisfy the Knill-Laflamme conditions \cite{PhysRevA.55.900}, meaning up to $\frac{N-1}{2}$ single-qubit phase flip errors are detectable, and provided additional ancilla qubits, correctable. 

Let us consider the traditional 3 qubit phase-flip error detection circuit depicted in Fig.\ref{fig:Shor3}.

\tikzset{
noisy/.style={starburst, line width=1pt,inner xsep=-4pt,inner ysep=-5pt, fill=yellow, draw=red}
}

\begin{figure}[htbp!]
    \begin{center}
    \begin{minipage}{\textwidth}
    \resizebox{\textwidth}{!}{
\begin{quantikz}[row sep={0.8cm,between origins}, wire types={q,q,q,q,q,n,n}]
\midstick[wires=3,brackets=right]{$\ket{\phi}$} & \gate{H} &[.25cm]  \gate[3,style={noisy, inner ysep=-28pt}]{\begin{matrix}\text{\color{red}single}\\ \text{\color{red} phase}\\\text{\color{red} flip}\end{matrix}} &[.25cm] \gate{H} \gategroup[5,steps=9,style={inner
xsep=2pt, fill=gray!10},background, label style={label
position=above, anchor=south, yshift=-.2cm}]{$U_\text{QEC}$} & \qw & \qw & \qw & \gategroup[5,steps=2,style={dashed,rounded
corners, inner
xsep=2pt, inner ysep=-7pt},background,label style={label
position=above,anchor=south, yshift=-.27cm}]{\text{\scriptsize parity check}}  \qw & \ctrl{0} & \qw & & \gate[3]{\text{\sc CORR}} & \midstick[wires=3,brackets=left]{$\ket{\phi}$}\\
 & \gate{H} &  & \gate{H} & \qw \gategroup[3,steps=2,style={dashed,rounded
corners, inner
xsep=2pt, inner ysep=-7pt},background,label style={label
position=above,anchor=south, yshift=-.27cm}]{\text{\scriptsize parity check}} & \ctrl{0} & \qw & \qw & \qw & \qw & \qw  & \qw  & \\
& \gate{H} &  & \gate{H} & \ctrl{0} & \qw & \qw & \ctrl{0} & \qw & \qw & \qw  & & \\[.5cm]
\midstick[wires=1,brackets=right]{$\ket{0}_a$} & \qw & \qw & \qw & \targ{}\vqw{-1} & \targ{}\vqw{-2} & \meter{} \\
\midstick[wires=1,brackets=right]{$\ket{0}_a$} & \qw & \qw & \qw & \qw & \qw & \qw & \targ{}\vqw{-2} & \targ{}\vqw{-4} & \meter{} \\[.5cm]
\midstick[wires=1]{$0$} & \cw & \cw & \cw & \cw & \cw & \cwbend{-2} & \cw & \cw & \cw & \cw\midstick[wires=2,]{$c$}&\cwbend{-3}&  \\
\midstick[wires=1]{$0$}& \cw & \cw & \cw & \cw & \cw & \cw & \cw & \cw & \cwbend{-2} & \cw&\cwbend{-1} &
\end{quantikz}
}
    \end{minipage}
    \caption{Traditional (Shor's) 3-qubit phase-flip error correction code. One can detect and correct $1$ phase flip error using 2 ancilla qubits for parity checks, and with initial state $\ket{\phi}=\alpha\ket{000}+\beta\ket{111}$. {\sc CORR} takes as an input the classical two-bits registry $c=(i)_2$, does nothing if $i=0$, and flips qubit $i$ otherwise, counting from top to bottom.}
    \label{fig:Shor3}
    \end{center}
\end{figure}
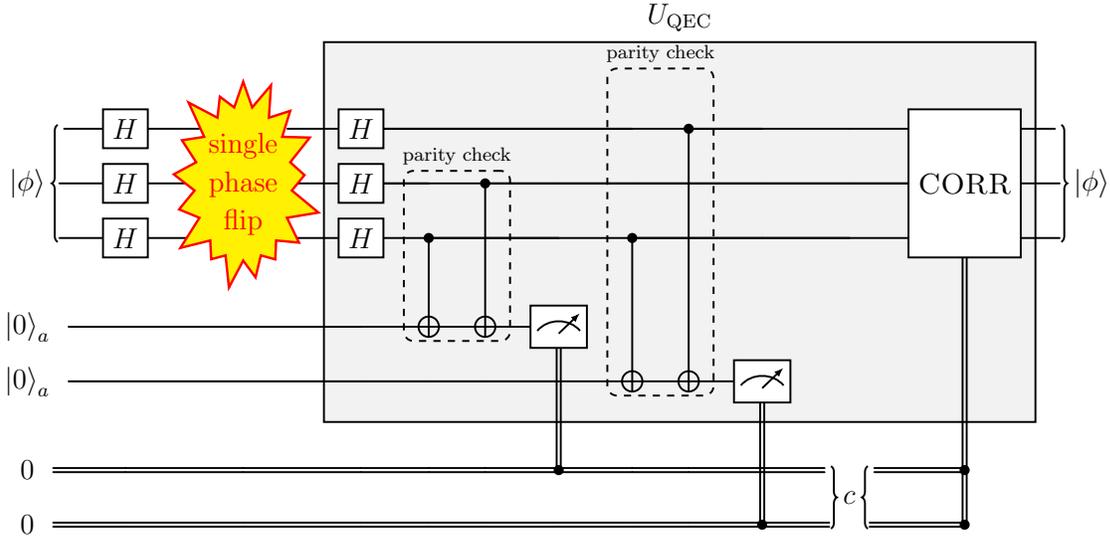

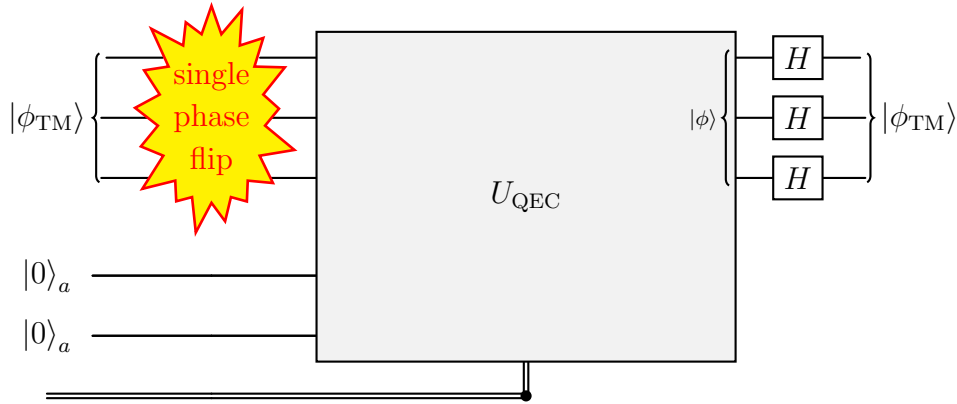
\begin{figure}[hbtp!]
    \centering
\begin{quantikz}[row sep={0.8cm,between origins}, wire types={q,q,q,q,q,n}]
\midstick[wires=3,brackets=right]{$\ket{\phi_\text{TM}}$} &[.25cm] \gate[3,style={noisy, inner ysep=-28pt}]{\begin{matrix}\text{\color{red}single}\\ \text{\color{red} phase}\\\text{\color{red} flip}\end{matrix}}   &[.25cm] \gate[5,style={fill=gray!10}]{\qquad\qquad\quad U_{\text{QEC}} \qquad\qquad\quad} \gateoutput[3]{$\ket{\phi}$} & \gate{H} & \midstick[wires=3,  brackets=left]{$\ket{\phi_\text{TM}}$}\\
  & & & \gate{H} &  \\
 & & &  \gate{H} & \\[.5cm]
\midstick[wires=1,brackets=right]{$\ket{0}_a$} & \qw & \qw \\
\midstick[wires=1,brackets=right]{$\ket{0}_a$} & \qw & \qw \\
\cw & \cw & \cwbend{-1}
\end{quantikz}
    \caption{3-qubit single phase-flip error correction code for one logical PTM state.}
    \label{fig:TMQEC}
\end{figure}

We notice, from (\ref{eq::HTM}), that in essence by applying the Hadamard gates at the beginning, the state was mapped to the PTM basis as $\ket{\phi_\text{TM}} = \frac{\alpha+\beta}{\sqrt{2}}\ket{0_\text{TM}^{(3)}}+\frac{\alpha-\beta}{\sqrt{2}}\ket{1_\text{TM}^{(3)}}$ just before the error step. Therefore, to adapt this error correction code to the PTM basis, it is only necessary to map $\ket{\phi}$ to $\ket{\phi_\text{TM}}$ by applying Hadamard gates at the end, as seen  on Fig.\ref{fig:TMQEC}. This provides an error-correcting gate that is applied after the error has occured, without the need for any state preparation prior to errors happening.

The same argument about error detection and correction can be applied to property 1.2.3 for errors due to a global magnetic field along the $z$ or $y$ direction.


\begin{property}Property 1.2.3 can be generalized for $d$-level quantum bases of information, or qu$d$its, if one defines the Thue-Morse states as:

\begin{equation}
     \ket{1_\text{TM}^{(N)}} = \sqrt{\frac{2}{d}} \sum_{k=0}^{d-1} t_k \ket{k}, \qquad 
     \ket{0_\text{TM}^{(N)}} = \sqrt{\frac{2}{d}} \sum_{k=0}^{d-1} \bar{t_k} \ket{k}.
\end{equation} 
\end{property}

Qudits may present certain advantages over qubits, providing a higher density of information, and a reduced number of non-local gates. Moreover their higher error rates, due to the increased number of excited states, can be compensated in certain platforms with fast gate times and slow decoherence times \cite{Jankovic2024, Hartmann2024}. In this case, it can be proven that PTM states provide noise detection for $J_z$ (spin-$\frac{d-1}{2}$ operator) up to order $\frac{1}{2}\left(\log_2(d)-1\right)$, providing an encoding of error robust states conceptually similar to Chiesa \textit{et al.} (2020) \cite{Chiesa2020}. 

\subsection{Noise-resistant quantum memories}

Property 1.2.1. implies that the PTM states are robust to external magnetic fields along any spatial direction to first order. 

Moreover, in the PTM logical basis, the different total spin operators take the following matrix forms:

\begin{equation}
    \ev{S_x}{0,1_\text{TM}^{(N)}} = \begin{pmatrix} 0 & 1\\ 1 & 0 \end{pmatrix}, \quad S_y = \begin{pmatrix} 0 & 0\\ 0 & 0 \end{pmatrix}, \quad S_z = \begin{pmatrix} 0 & 0\\ 0 & 0 \end{pmatrix}.
\end{equation}

This makes the PTM states particularly useful for storing a logical qubit in an environment with strong magnetic noise along the $y$ or $z$-direction. This is not applicable in the $x$-direction, as only the states $\ket{0_\text{TM}^{(N)}}$ and $\ket{1_\text{TM}^{(N)}}$ are robust, but not superposition states.

\subsection{Link to quantum chaos}

From the analysis presented in (\ref{eq::HTM}), it becomes evident that the last row of the Walsh-Hadamard matrix corresponds to a modified PTM sequence, specifically represented as $(-1)^{t_n}$. This particular element and its connection to quantum chaos has been explored in depth in \cite{PhysRevE.71.065303,PhysRevE.74.035203}, where the interplay between the PTM sequence and the quantum Fourier transform is investigated. This synergy is central to the construction of an approximate multifractal eigenstate of the quantum baker's map, providing a novel approach to unraveling the complexity of quantum chaotic systems. Such findings not only highlight the multifaceted role of the PTM sequence within quantum chaos but also contribute significantly to our broader understanding of the dynamical properties of quantum systems.
\todo*{graph pour illustrer}

\begin{definition}
An example of these interesting properties can be observed by defining the $N$-qubits Quantum Fourier Transform {\sc $QFT$($N$)} gate, elementwise ($0 \leq j,k < N$)
\begin{equation}
    \left(QFT(N)\right)_{jk} = \frac{1}{\sqrt{2^N}}e^{i\frac{2\pi}{2^N}jk}.
\end{equation}

This gate, that applies the discrete Fourier transform over the Hilbert space of the qubits, is essential for some quantum algorithms such as Shor's factorization algorithm \cite{Shor1997}.
\end{definition}

Applying the $QFT$ gate to a state in the PTM basis proves interesting, since

\begin{equation}
    \text{$QFT$($N$)}\left(\alpha \ket{0_\text{TM}^{(N)}} + \beta \ket{1_\text{TM}^{(N)}}\right) = \frac{\sqrt{2}}{2^{N}}\sum_{j=0}^{2^N-1} \sum_{k=0}^{2^N-1} \frac{1}{2} \left(\alpha + \beta + (-1)^{t_k}(\alpha-\beta)\right) e^{i\frac{2\pi}{2^N}jk} \ket{(j)_2}
\end{equation}

Therefore, by defining $F^{TM}_j \equiv \mel{j}{\text{$QFT$($N$)}\Big(\alpha \ket{0_\text{TM}^{(N)}} + \beta}{1_\text{TM}^{(N)}}\Big)$ and
since $\forall j \in \mathbb{N}^*,\ \sum_{k=0}^{2^N-1} e^{i\frac{2\pi}{2^N}jk} = 0$, we have
\begin{equation}
   F^{TM}_{j\neq0}  = \frac{\alpha-\beta}{2^{N}\sqrt{2}} \sum_{k=0}^{2^N-1} (-1)^{t_k} e^{i\frac{2\pi}{2^N}jk}.
\end{equation}

Moreover \cite{TMmaths}, \begin{equation}\forall x\in \mathbb{R},\ \prod_{k=0}^{N-1} \left(1 - x^{2^k}\right) = \sum_{k=0}^{2^N-1} (-1)^{t_k} x^k, \end{equation} 

hence, \begin{equation}
   F^{TM}_{j\neq0}   = \frac{\alpha-\beta}{2^{N}\sqrt{2}}  \prod_{k=0}^{N-1} \left(1 - e^{i\frac{2\pi j}{2^{N-k}}}\right).
\end{equation}

The specific case of $j=0$, is given by
\begin{equation}
    F^{TM}_{0} = \frac{\alpha+\beta}{\sqrt{2}}.
\end{equation}

Further simplifying, we obtain, for $j\neq 0$

\begin{equation}
    \left|F^{TM}_j\right|^2 = \frac{(\alpha-\beta)^2}{2}\prod_{k=0}^{N-1}\sin^2\left(\pi j 2^{k-N}\right).
\end{equation}

\begin{figure}[htbp!]
    \centering
    \includegraphics[width=\textwidth]{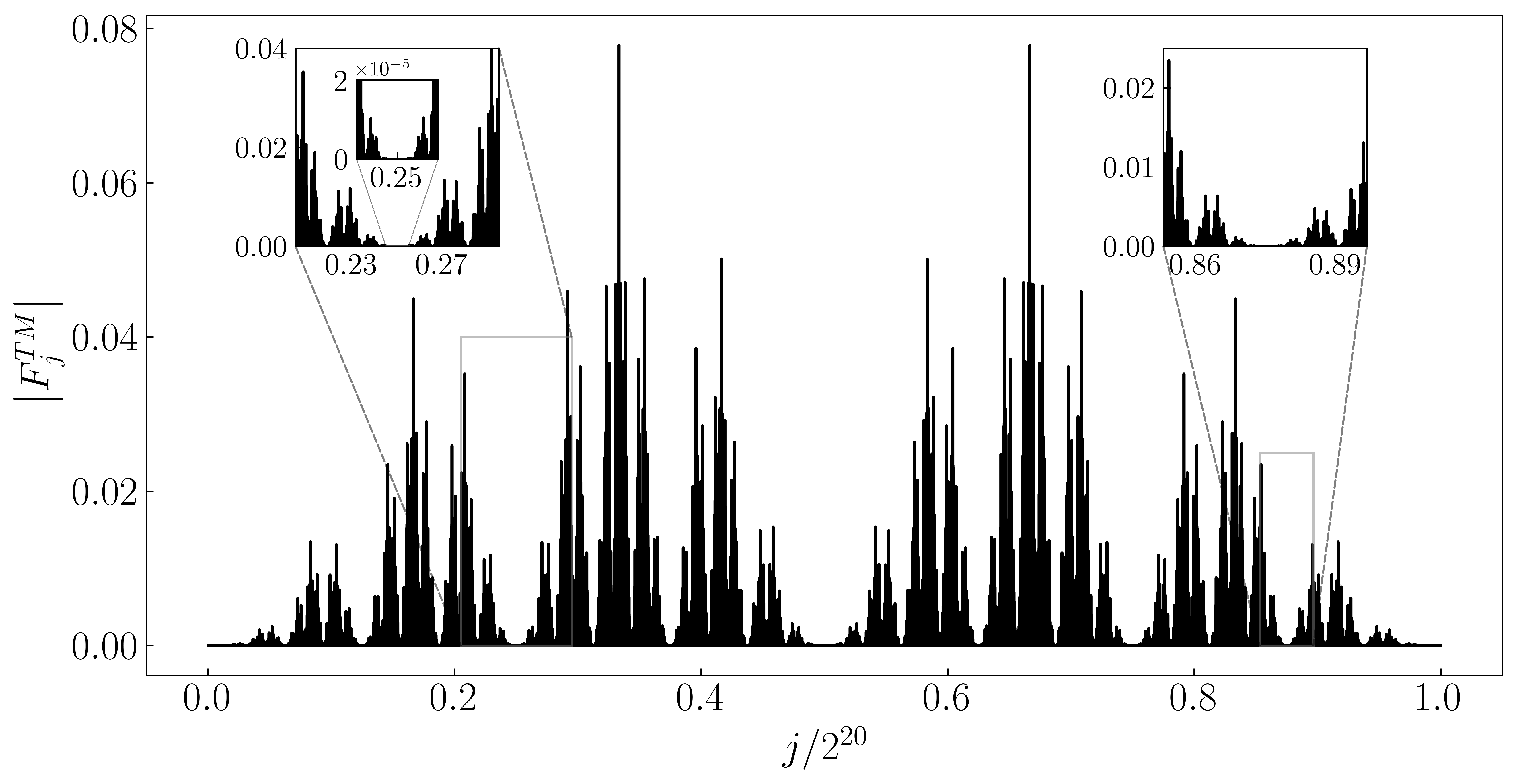}
    \caption{Amplitudes of $F^{TM}_j = \frac{1}{\sqrt{2}}\mel{j}{QFT(20)\Big( \ket{0_\text{TM}^{(20)}} - }{1_\text{TM}^{(20)}}\Big) = \mel{j}{QFT(20)\otimes_{k=1}^{20} H^{(k)}}{(2^{20}-1)_2}$ for $j=0$ to $2^{20}-1$. Two zooms, around 0.2 and 0.88 show the self-similarity of the curve. An additional zoom around 0.25 further shows the self similar behaviour.}
    \label{fig:fractal}
\end{figure}

For the case $\alpha = -\beta = \frac{1}{\sqrt{2}}$, $\left|F^{TM}_j\right|^2 = \mel{j}{QFT(N)\otimes_{k=1}^N H^{(k)}}{(2^N-1)_2}$ are shown in Fig.\ref{fig:fractal} for $N=20$. One can notice self-similarity and infer the multifractal nature of the state in the Hilbert space.

It is these multifractal properties \cite{multifrac} and the recursive definition of the Thue-Morse states that have been used in to construct, for example, approximate eigenstates of the quantum baker map $B(N)$ \cite{PhysRevE.71.065303}.

\begin{equation}
    B(N) = QFT(N)^{-1} \begin{pmatrix}
        QFT(N-1) & 0 \\ 0 & QFT(N-1)
    \end{pmatrix}.
\end{equation}

Moreover, considering the use of both the Hadamard gate $\otimes_{k=1}^N H^{(k)}$ and QFT gate $QFT(N)$ in Shor's factorization algorithm, it would be interesting to study further the appearence of quantum chaos that we showed to be arising from their interactions in such a well-known quantum algorithm \cite{PhysRevE.74.035203}.

\subsection{Link with number theory}

There are several interesting formulae that use the PTM sequence \cite{Allouche1999,Toth}, including:

\begin{itemize}
    \item Its product generating function formula 
    \begin{equation}\forall x \in \mathbb{R}, \quad \prod_{i=0}^{N} \left(1 - x^{2^i}\right) = \sum_{j=0}^{2^{N+1}-1} (-1)^{t_j} x^j,
    \end{equation} it can also be generalized for $N\rightarrow\infty$, \begin{equation}\forall x \in \mathbb{R}, \quad \prod_{i=0}^{\infty} \left(1 - x^{2^i}\right) = \sum_{j=0}^{\infty} (-1)^{t_j} x^j\end{equation}. 
    \item$\begin{aligned}\prod_{n=0}^{\infty}\left(\frac{2 n+1}{2 n+2}\right)^{2\left(1-t_n\right)}\left(\frac{2 n+3}{2 n+2}\right)=\frac{\sqrt{2}}{\pi}\end{aligned}$
    \item$\begin{aligned}\prod_{n=0}^{\infty}\left(\frac{2 n+1}{2 n+2}\right)^{2 t_n}\left(\frac{2 n+3}{2 n+2}\right)=\frac{2 \sqrt{2}}{\pi}\end{aligned}$
    \item$\begin{aligned}\prod_{n=0}^{\infty}\left(\frac{2 n+1}{2 n+2}\right)^{(-1)^{t_n}}=\frac{\sqrt{2}}{2}\end{aligned}$
    \item$\begin{aligned}\prod_{n=1}^{\infty}\left(\frac{2 n}{2 n+1}\right)^{(-1)^{t_n}} = 1.6281\dots\end{aligned}$ converges towards a number whose algebraic nature remains unknown \cite{Allouche1999}.
    \item $\begin{aligned}\tau = \sum_{i=0}^\infty \frac{t_i}{2^{i+1}} = 0.41245\dots\end{aligned}$ is known as the PTM constant and has been proved to be a transcendental number. 
    \item $\begin{aligned}\forall s \in \mathbb{C} \text{ s.t. }\text{Re}(s)>1,\quad  \zeta(s) = (1+\frac{1}{2^s}) \sum_{n\geq1} \frac{t_{n-1}}{n^s} + (1-\frac{1}{2^s}) \sum_{n\geq1} \frac{t_{n}}{n^s},\end{aligned}$ \todo*{convergence en $t$?} where $\zeta$ is the Riemann Zeta function. 
\end{itemize}

In particular, the last formula presents an interesting linear combination of Dirichlet series. When linked to the work of Feiler and Schleich \cite{Feiler2015}, it becomes of particular interest as it provides a link between quantum computing and number theory.  Just as in the cited work, one can obtain the result of this sum of Dirichlet series using the interference of two initial quantum states in a non-linear interaction with logarithmic energy spectrum. Let us assume a given system evolves under a non-linear oscillator Hamiltonian with a logarithmic energy spectrum: $H=\hbar\omega\ln(n+1)\ket{n}\bra{n}$ \cite{Gleisberg2013}. Given a certain $s=\sigma+i\tau$ complex argument, one can consider two non-normalized states:
\begin{equation}
    \begin{aligned}
        \ket{\psi_1} &= \sum_{n\ge0} t_{n}\cdot(n+1)^{-\sigma/2}\ket{n}\\
        \ket{\psi_2} &= \sum_{n\ge0} t_{n+1}\cdot(n+1)^{-\sigma/2}\ket{n}.
    \end{aligned}
\end{equation}

If one measures the conjugate auto-correlation probability amplitude $\bra{\psi^*}\ket{\psi(t)}$ for $\ket{\psi_1}$ (resp. $\ket{\psi_2}$), evolved for time $t=\tau/\omega$, the result will be proportional to $\sum_{n\geq1} \frac{t_{n-1}}{n^s}$ (resp. $\sum_{n\geq1} \frac{t_{n}}{n^s}$). 

Following the arguments of \textit{Feiler et al. }\cite{Feiler2015}, we remark that by evolving the initial state, normalized by $\mathcal{N}$,
\begin{equation}
    \ket{\psi_0} = \mathcal{N} \left((1+\frac{1}{2^s})\ket{\psi_1} + (1-\frac{1}{2^s})\ket{\psi_2}\right),
\end{equation}
for a time \( t=\tau/g \) under an interaction with a logarithmic energy spectrum in the interaction picture \cite{Feiler2013} ($H_I=\hbar g \ln(\hat{n}+1)\hat{\sigma_z}$), the measurement of the conjugate auto-correlation probability amplitude $\bra{\psi_0^*}\ket{\psi_0(t=\tau/g)}$ is expected to yield the exact value of $\zeta(s) = \mathcal{N}^2 \left[ (1+\frac{1}{2^s}) \sum_{n\geq1} \frac{t_{n-1}}{n^s} + (1-\frac{1}{2^s}) \sum_{n\geq1} \frac{t_{n}}{n^s} \right].$

This provides a method to determine the value of the Riemann Zeta function through quantum measurements.

\section{Conclusion}

In this study, we have explored the many manifestations and uses of the Prouhet-Thue-Morse (PTM) sequence in quantum computing, highlighting its intrinsic mathematical interest and practical significance. Through rigorous analysis, we have shown that the PTM sequence, beyond its mathematical appeal, plays a central role in quantum error correction, the design of noise-resistant quantum memories. We also elucidated its links with quantum chaos and number theory.

Our investigations have shown how logical states whose encoding is based on the PTM sequence exhibit remarkable properties, such as resilience to spin flip errors and low sensitivity to external magnetic fields, thus providing valuable insights for the design of fault-tolerant quantum computing architectures. Furthermore, the integration of the sequence into the Hilbert space of quantum computing platforms, particularly in the context of the $X$-$X$ Ising chain and spin systems, demonstrates its potential to increase the efficiency and reliability of quantum information processing.

The connection between the PTM sequence and quantum chaos, through the Walsh-Hadamard transform and the quantum baker’s map, enriches our understanding of the dynamical behaviors of quantum systems and provides a novel lens through which to view quantum chaos. Moreover, the intriguing links between the PTM sequence and number theory, exemplified by its relation to the Riemann zeta function and Dirichlet series, open new avenues for interdisciplinary research, bridging quantum computing and mathematical number theory.

While the novelty of our results is tempered by the fundamental nature of the PTM sequence in both mathematics and physics, the implications of our findings for quantum computing are profound. The insights gained from this study not only contribute to the ongoing development of quantum computing technologies but also stimulate further research into the underlying mathematical structures that govern quantum mechanics.

As we continue to unravel the complexities of quantum systems and their computing capabilities, the PTM sequence stands as a testament to the deep connections between abstract mathematical concepts and practical quantum technologies. Our exploration of these links also highlights the importance of interdisciplinary research, in this case, physics and mathematics, in order to reveal the full potential of quantum computing.

\section{Acknoledgments}

This work was funded by the French National Research Agency (ANR) through the Programme d'Investissement d'Avenir under contract ANR-11-LABX-0058\_NIE and ANR-17-EURE-0024 within the Investissement d’Avenir program ANR-10-IDEX-0002-02. The authors would like to thank the Online Encyclopedia of Integer Sequences (OEIS) for providing valuable data and resources that contributed to this research. Specifically, the (PTM) sequence A010060.

\section{Data availability}

The codes and calculations done in this study are available from the authors upon reasonable request.

\section{Conflicts of interest}

The authors declare no conflicts of interests.

\section{References}
\bibliographystyle{iopart-num}
\bibliography{references}

\providecommand{\newblock}{}
\begin{thebibliography}{10}
\expandafter\ifx\csname url\endcsname\relax
  \def\url#1{{\tt #1}}\fi
\expandafter\ifx\csname urlprefix\endcsname\relax\def\urlprefix{URL }\fi
\providecommand{\eprint}[2][]{\url{#2}}

\bibitem{Prouhet1851}
Prouhet E 1851 {\em C. R. Acad. Sci. Paris Ser. I\/} {\bf 33} 225

\bibitem{Allouche1999}
Allouche J~P and Shallit J 1999 The ubiquitous prouhet-thue-morse sequence {\em Sequences and their Applications\/} ed Ding C, Helleseth T and Niederreiter H (London: Springer London) pp 1--16 ISBN 978-1-4471-0551-0

\bibitem{PhysRevB.43.1034}
Kol\'a\ifmmode~\check{r}\else \v{r}\fi{} M, Ali M~K and Nori F 1991 {\em Phys. Rev. B\/} {\bf 43}(1) 1034--1047 \urlprefix\url{https://link.aps.org/doi/10.1103/PhysRevB.43.1034}

\bibitem{Thue1906}
Thue A 1906 {\em Kra. Vidensk. Selsk. Skrifter. I. Mat.-Nat. Kl.\/}

\bibitem{MorseRecurrentGO}
Morse H~M {\em Transactions of the American Mathematical Society\/} {\bf 22} 84--100 \urlprefix\url{https://api.semanticscholar.org/CorpusID:54043171}

\bibitem{xiong2022topological}
Xiong L, Zhang Y, Liu Y, Zheng Y and Jiang X 2022 {\em Physical Review Applied\/} {\bf 18} 064089 \urlprefix\url{https://link.aps.org/doi/10.1103/PhysRevApplied.18.064089}

\bibitem{DENG2011360}
Deng X~H, ji~Ren~Yuan, Hong W~Q and Ouyang H 2011 {\em Physics Procedia\/} {\bf 22} 360--365 ISSN 1875-3892 2011 International Conference on Physics Science and Technology (ICPST 2011) \urlprefix\url{https://www.sciencedirect.com/science/article/pii/S1875389211007103}

\bibitem{Matarazzo2010}
Matarazzo V, De~Nicola S, Zito G, Mormile P, Rippa M, Abbate G, Zhou J and Petti L 2010 {\em Journal of Optics\/} {\bf 13} 015602 ISSN 2040-8986 \urlprefix\url{http://dx.doi.org/10.1088/2040-8978/13/1/015602}

\bibitem{6326756}
Yang J~K, Noh H, Boriskina S~V, Rooks M~J, Solomon G~S, Dal~Negro L and Cao H 2012 Lasing in thue-morse structure with optimal aperiodicity {\em 2012 Conference on Lasers and Electro-Optics (CLEO)\/} pp 1--2

\bibitem{TMmaths}
Nguyen H~D 2014 A new proof of the prouhet-tarry-escott problem \urlprefix\url{https://arxiv.org/abs/1411.6168}

\bibitem{Aifer2022}
Aifer M and Deffner S 2022 {\em New Journal of Physics\/} {\bf 24} 055002 ISSN 1367-2630 \urlprefix\url{http://dx.doi.org/10.1088/1367-2630/ac6821}

\bibitem{PhysRevA.55.900}
Knill E and Laflamme R 1997 {\em Phys. Rev. A\/} {\bf 55}(2) 900--911 \urlprefix\url{https://link.aps.org/doi/10.1103/PhysRevA.55.900}

\bibitem{Jankovic2024}
Janković D, Hartmann J~G, Ruben M and Hervieux P~A 2024 {\em npj Quantum Information\/} {\bf 10} ISSN 2056-6387 \urlprefix\url{http://dx.doi.org/10.1038/s41534-024-00829-6}

\bibitem{Hartmann2024}
Hartmann J~G, Janković D, Pasquier R, Ruben M and Hervieux P~A 2024 Nonlinearity of the fidelity in open qudit systems: Gate and noise dependence in high-dimensional quantum computing \urlprefix\url{https://arxiv.org/abs/2406.15141}

\bibitem{Chiesa2020}
Chiesa A, Macaluso E, Petiziol F, Wimberger S, Santini P and Carretta S 2020 {\em The Journal of Physical Chemistry Letters\/} {\bf 11} 8610–8615 ISSN 1948-7185 \urlprefix\url{http://dx.doi.org/10.1021/acs.jpclett.0c02213}

\bibitem{PhysRevE.71.065303}
Meenakshisundaram N and Lakshminarayan A 2005 {\em Phys. Rev. E\/} {\bf 71}(6) 065303 \urlprefix\url{https://link.aps.org/doi/10.1103/PhysRevE.71.065303}

\bibitem{PhysRevE.74.035203}
Maity K and Lakshminarayan A 2006 {\em Phys. Rev. E\/} {\bf 74}(3) 035203 \urlprefix\url{https://link.aps.org/doi/10.1103/PhysRevE.74.035203}

\bibitem{Shor1997}
Shor P~W 1997 {\em SIAM Journal on Computing\/} {\bf 26} 1484–1509 ISSN 1095-7111 \urlprefix\url{http://dx.doi.org/10.1137/S0097539795293172}

\bibitem{multifrac}
Fan A, Schmeling J and Shen W 2022 Multifractal analysis of generalized thue-morse trigonometric polynomials \urlprefix\url{https://arxiv.org/abs/2212.13234}

\bibitem{Toth}
Tóth L 2022  \urlprefix\url{https://arxiv.org/abs/2211.13570}

\bibitem{Feiler2015}
Feiler C and Schleich W~P 2015 {\em New Journal of Physics\/} {\bf 17} 063040 ISSN 1367-2630 \urlprefix\url{http://dx.doi.org/10.1088/1367-2630/17/6/063040}

\bibitem{Gleisberg2013}
Gleisberg F, Mack R, Vogel K and Schleich W~P 2013 {\em New Journal of Physics\/} {\bf 15} 023037 ISSN 1367-2630 \urlprefix\url{http://dx.doi.org/10.1088/1367-2630/15/2/023037}

\bibitem{Feiler2013}
Feiler C and Schleich W~P 2013 {\em New Journal of Physics\/} {\bf 15} 063009 ISSN 1367-2630 \urlprefix\url{http://dx.doi.org/10.1088/1367-2630/15/6/063009}

\end{thebibliography}

\appendix

\section{Notations}
\label{apdx:def}

\begin{definition}[Hilbert Space]
A \emph{Hilbert Space} is a finite-dimensional (\(d\)) vector space over \(\mathbb{C}\), equipped with a sesquilinear inner product \(\langle \cdot | \cdot \rangle\). The norm of a vector \(|\psi\rangle \in \mathcal{H}\) is defined as:
\[
\|\psi\| = \sqrt{\langle \psi | \psi \rangle}.
\]
In this work, we consider \(\mathcal{H} = \mathbb{C}^d\), where \(d\) is the dimension of the space.
\end{definition}

\begin{definition}[Basis]
An \emph{orthonormal basis} of a Hilbert space \(\mathcal{H}\) of dimension $d$ is a set of vectors \(\{|i\rangle\}_{i=1}^d\) such that:
\[
\langle i | j \rangle = \delta_{ij}, \quad \forall \, i, j,
\]
where \(\delta_{ij}\) is the Kronecker delta. This basis is often referred to as the computational basis.
\end{definition}

\begin{definition}[$\ket{\cdot}$, Ket Notation]
Let \(\mathcal{H}\) be a finite-dimensional Hilbert space over \(\mathbb{C}\). A \emph{ket} is a vector \(|\psi\rangle \in \mathcal{H}\), where \(|\psi\rangle\) denotes an abstract element of the vector space. Kets are typically used to represent quantum states.
\end{definition}

\begin{definition}[$\bra{\cdot}$, Bra Notation]
Let \(\mathcal{H}\) be a Hilbert space, and let \(\mathcal{H}^*\) denote its dual space, which is the space of all linear functionals (linear maps from \(\mathcal{H}\) to \(\mathbb{C}\)). A \emph{bra} is an element of the dual space, denoted as \(\langle \phi|\), and acts on kets via the inner product:
\[
\langle \phi|: \mathcal{H} \to \mathbb{C}, \quad \langle \phi| \psi \rangle = \text{inner product of } |\phi\rangle \text{ and } |\psi\rangle.
\]
Through the seminal Riesz representation theorem, the mapping \(|\psi\rangle \mapsto \langle \psi|\) is a conjugate-linear isomorphism defined by:
\[
\langle \psi| \phi \rangle = \overline{\langle \phi| \psi \rangle}.
\]
\end{definition}

\begin{definition}[$\braket{\cdot}{\cdot}$, Dirac Notation (Bra-Ket Notation)]
The inner product of two state vectors \(|\psi\rangle, |\phi\rangle \in \mathcal{H}\) is denoted by:
\[
\langle \psi | \phi \rangle \in \mathbb{C}.
\]
For a ket \(|\psi\rangle\), the outer product is an operator \(|\psi\rangle \langle \psi|\) that acts on the Hilbert space:
\[
|\psi\rangle \langle \psi| : \mathcal{H} \to \mathcal{H}.
\]
The bra-ket notation provides a convenient way to express quantum mechanical operations and relationships.
\end{definition}

\begin{remark}[Orthogonality and Orthonormality]
Two states \(|\psi\rangle\) and \(|\phi\rangle\) are orthogonal if \(\langle \psi | \phi \rangle = 0\). A basis \(\{|i\rangle\}\) is orthonormal if:
\[
\langle i | j \rangle = \delta_{ij}, \quad \forall i, j.
\]
\end{remark}

\begin{definition}[State Vectors]
A \emph{state vector} in the Hilbert space is a normalized vector \(|\psi(t)\rangle \in \mathcal{H}\) expressed in the computational basis as:
\[
|\psi(t)\rangle = \sum_{i=1}^d \alpha_i(t) |i\rangle,
\]
where the coefficients \(\alpha_i(t) \in \mathbb{C}\) satisfy the normalization condition:
\[
\sum_{i=1}^d |\alpha_i(t)|^2 = 1.
\]
\end{definition}

For example,for a single qubit (\(d = 2\)), the computational basis is \(\{|0\rangle, |1\rangle\}\). A general state vector is written as:
\[
|\psi\rangle = \alpha_0 |0\rangle + \alpha_1 |1\rangle, \quad \text{where } |\alpha_0|^2 + |\alpha_1|^2 = 1.
\]

\begin{definition}[$\otimes$, Tensor Product and composite systems]
Suppose \(V_1\) and \(V_2\) are vector spaces over a field \(\mathbb{F}\). A \emph{tensor product} of \(V_1\) and \(V_2\) is a vector space \(W\) over \(\mathbb{F}\) together with a bilinear map \(T : V_1 \otimes V_2 \to W\) having the following \emph{universal property}:

If \(U\) is any vector space over \(\mathbb{F}\) and \(\Phi : V_1 \otimes V_2 \to U\) is a bilinear map, then there exists a unique linear map \(\hat{\Phi} : W \to U\) such that the following diagram commutes:
\[
\begin{array}{ccc}
V_1 \otimes V_2 & \xrightarrow{T} & W \\
\downarrow{\Phi} & & \downarrow{\hat{\Phi}} \\
U & & U
\end{array}
\]

In a quantum system consisting of \(n\) subsystems, the total Hilbert space is the tensor product of the individual Hilbert spaces:
\[
\mathcal{H}_\text{total} = \mathcal{H}_1 \otimes \mathcal{H}_2 \otimes \cdots \otimes \mathcal{H}_n.
\]
If each subsystem has dimension \(d\), then the total Hilbert space has dimension \(d^n\).

For state vectors \(\ket{a} \in \mathcal{H}_1\) and \(\ket{b} \in \mathcal{H}_2\), the tensor product is defined as:
\[
\ket{a} \otimes \ket{b} \in \mathcal{H}_\text{total}.
\]
This is often written in shorthand as:
\[
\ket{a} \otimes \ket{b} = \ket{a} \ket{b} = \ket{ab}.
\]

For a general multipartite state, we write:
\[
\ket{\psi} = \ket{\psi_1} \otimes \ket{\psi_2} \otimes \cdots \otimes \ket{\psi_n},
\]
or in shorthand:
\[
\ket{\psi} = \ket{\psi_1 \psi_2 \cdots \psi_n}.
\]
\end{definition}

\begin{remark}[Tensor Product of Basis States]
The computational basis of the composite system is the tensor product of the individual bases. For \(n\) subsystems of dimension $d$, the computational basis is:
\[
\big\{ \ket{i_1} \otimes \ket{i_2} \otimes \cdots \otimes \ket{i_n} \big\},
\]
or in shorthand:
\[
\big\{ \ket{i_1 i_2 \cdots i_n} \big\},
\]
where \(i_k \in \{1, 2, \ldots, d\}\) for each subsystem.
\end{remark}

For example, if \(d = 2\) (qubits), the computational basis is:
\[
\big\{ \ket{00}, \ket{01}, \ket{10}, \ket{11} \big\}.
\]

\begin{definition}[Density Matrix]
The density matrix for a composite system is defined on the total Hilbert space \(\mathcal{H}_\text{total}\). For a pure state \(\ket{\psi} \in \mathcal{H}_\text{total}\), it is:
\[
\rho = \ket{\psi} \bra{\psi}.
\]
For a general mixed state, \(\rho\) is a positive semi-definite operator with unit trace on \(\mathcal{H}_\text{total}\).
\end{definition}

\begin{definition}[$\lbrack \cdot, \cdot \rbrack$, Commutator]
Let \(A\) and \(B\) be two operators acting on a vector space or a Hilbert space. The \emph{commutator} of \(A\) and \(B\), denoted by \([A, B]\), is defined as:
\[
[A, B] = AB - BA.
\]
\end{definition}

\begin{definition}[$\{\cdot,\cdot\}$, Anticommutator]
Let \(A\) and \(B\) be two operators acting on a vector space or a Hilbert space. The \emph{anticommutator} of \(A\) and \(B\), denoted by \(\{A, B\}\), is defined as:
\[
\{A, B\} = AB + BA.
\]
\end{definition}

\section{Proofs of properties}

\subsection{Proof of Property 1.2.1} $$\forall N>1,  \ev**{\sum_{k=1}^N \sigma_{x,y,z}^{(k)}}{0_\text{TM}^{(N)}} = 0\text{ \textit{and} }\ev**{\sum_{k=1}^N \sigma_{x,y,z}^{(k)}}{1_\text{TM}^{(N)}} = 0$$

\textbf{Proof for $\sigma_z$:} 
(by induction)
\begin{itemize}
    \item One can easily check that for $N=2$ it is true. Using $\sigma^{(1)}_z + \sigma^{(2)}_z = \sigma_z \otimes \mathbb{1}_2 +  \mathbb{1}_2 \otimes \sigma_z.$
    $$\begin{aligned}\ev**{\left(\sigma^{(1)}_z + \sigma^{(2)}_z\right)}{0_{TM}^{(2)}} &=
    \begin{pmatrix}
        1&0&0&1
    \end{pmatrix} \begin{pmatrix}
        2 & 0 & 0 & 0\\
        0 & 0 & 0 & 0\\
        0 & 0 & 0 & 0\\
        0 & 0 & 0 & -2
    \end{pmatrix}
    \begin{pmatrix}
        1\\0\\0\\1
    \end{pmatrix} = 2-2 = 0,\\
    \ev**{\left(\sigma^{(1)}_z + \sigma^{(2)}_z\right)}{1_{TM}^{(2)}} &=
    \begin{pmatrix}
        0&1&1&0
    \end{pmatrix} \begin{pmatrix}
        2 & 0 & 0 & 0\\
        0 & 0 & 0 & 0\\
        0 & 0 & 0 & 0\\
        0 & 0 & 0 & -2
    \end{pmatrix}
    \begin{pmatrix}
        0\\1\\1\\0
    \end{pmatrix} = 0.\end{aligned}$$

    \item Using the recursive properties of the Thue-Morse sequence, $$\begin{aligned}&\ev**{\sum_{k=1}^N \sigma_{z}^{(k)}}{0_{TM}^{(N)}}\\&=\ev**{\ev**{\sum_{k=1}^N \sigma_{z}^{(k)}}{0}}{0_{TM}^{(N-1)}} + \mel**{0_{TM}^{(N-1)}}{\mel**{0}{\sum_{k=1}^N \sigma_{z}^{(k)}}{1}}{1_{TM}^{(N-1)}}\\&+\mel**{1_{TM}^{(N-1)}}{\mel**{1}{\sum_{k=1}^N \sigma_{z}^{(k)}}{0}}{0_{TM}^{(N-1)}} + \ev**{\ev**{\sum_{k=1}^N \sigma_{z}^{(k)}}{1}}{1_{TM}^{(N-1)}}\end{aligned}$$
    and
    $$\begin{aligned}&\ev**{\sum_{k=1}^N \sigma_{z}^{(k)}}{1_{TM}^{(N)}}\\&=\ev**{\ev**{\sum_{k=1}^N \sigma_{z}^{(k)}}{0}}{1_{TM}^{(N-1)}} + \mel**{1_{TM}^{(N-1)}}{\mel**{0}{\sum_{k=1}^N \sigma_{z}^{(k)}}{1}}{0_{TM}^{(N-1)}}\\&+\mel**{0_{TM}^{(N-1)}}{\mel**{1}{\sum_{k=1}^N \sigma_{z}^{(k)}}{0}}{1_{TM}^{(N-1)}} + \ev**{\ev**{\sum_{k=1}^N \sigma_{z}^{(k)}}{1}}{0_{TM}^{(N-1)}}.\end{aligned}$$
    Assuming the property holds true for $N-1$, and since $\sigma_{z}^{(k)}$ is diagonal, both equations reduce to $\sigma_z^{(1)}$ acting on the added qubit and are equal to:
    $$\ev*{\sigma_z}{0} + \ev*{\sigma_z}{1} = 1-1 = 0$$
\end{itemize}

\textbf{Proof for $\sigma_x$ and $\sigma_y$:} Since the effect of $\sigma_x^{(k)}$ is to flip the $k^\text{th}$ qubit, $\sum_{k=1}^N\sigma_x^{(k)}$ maps $\ket*{0_{TM}^{(N)}} \leftrightarrow \ket*{1_{TM}^{(N)}}$ which are orthogonal to each other, hence the property is true.\\
For $\sigma_y$, it follows from $\sigma_y = -i\sigma_z\sigma_x$. Therefore, by defining $c_k:=0,1$ as the state of the $k^\text{th}$ qubit and $\overline{c_k}=1-c_k$ $$\begin{aligned}\ev**{\sum_{k=1}^N \sigma_y^{(k)}}{0,1_{TM}^{(N)}} &= -i \ev**{\sum_{k=1}^N \sigma_z^{(k)}\sigma_x^{(k)}}{0,1_{TM}^{(N)}} =  -i \sum_{k=1}^N \ev**{\sigma_z^{(k)}\sigma_x^{(k)}}{0,1_{TM}^{(N)}} \\
&=  -i \sum_{k=1}^N \ev**{\sigma_z^{(k)}\sigma_x^{(k)}}{0,1_{TM}^{(N)}} =   -i \sum_{k=1}^N \mel**{c_k}{\sigma_z}{\overline{c_k}} = 0. \end{aligned}$$

\subsection{Proof of Property 1.2.2}

\textit{Let $Z$ be an ensemble of $\sigma_z^{(k)}$ acting on $M$ qubits ($M<N$) \\ $$\ev**{\prod_{k\in Z} \sigma_z^{(k)}}{0_\text{TM}^{(N)}} = \ev**{\prod_{k\in Z} \sigma_z^{(k)}}{1_\text{TM}^{(N)}},$$ $$\text{and }\mel**{1_\text{TM}^{(N)}}{\prod_{k\in Z} \sigma_z^{(k)}}{0_\text{TM}^{(N)}} = 0.$$}

 First, we recall that $\sigma_z$ is involutory i.e. $({\sigma_z^{(k)}})^2 = \mathbb{1}$ and having a $\sigma_z^{(k)}$ appear an odd amount of time is equivalent to it appearing only once. Therefore, we notice $$ \prod_{k\in Z} \sigma_z^{(k)} = \bigotimes_{k\in Z'} \sigma_z^{(k)} $$ with $Z'$ being the ensemble of only one occurrence of the $\sigma^{(k)}$s that appear an odd amount of times and not the others.
\\Induction:\\
\begin{itemize}
    \item One can check that for $N=2$, the equality does not hold for $Z' = (\sigma_z^{(1)}, \sigma_z^{(2)})$, but does for any subset.
    $$ \begin{aligned} &\left.\begin{cases} 
    \ev**{\sigma_z^{(1)} \otimes \mathbb{1}_2}{0_{TM}^{(2)}} &= \begin{pmatrix}
        1&0&0&1
    \end{pmatrix} \begin{pmatrix}
        1 & 0 & 0 & 0\\
        0 & 1 & 0 & 0\\
        0 & 0 & -1 & 0\\
        0 & 0 & 0 & -1
    \end{pmatrix}
    \begin{pmatrix}
        1\\0\\0\\1
    \end{pmatrix} = 1-1 = 0 \\
    \ev**{\sigma_z^{(1)} \otimes \mathbb{1}_2}{1_{TM}^{(2)}} &= 
    \begin{pmatrix}
        0&1&1&0
    \end{pmatrix} \begin{pmatrix}
        1 & 0 & 0 & 0\\
        0 & 1 & 0 & 0\\
        0 & 0 & -1 & 0\\
        0 & 0 & 0 & -1
    \end{pmatrix}
    \begin{pmatrix}
        0\\1\\1\\0
    \end{pmatrix} = 1-1 = 0  
    \end{cases}\right.
    \\
    &\left.\begin{cases} 
    \ev**{\mathbb{1}_2\otimes \sigma_z^{(2)}}{0_{TM}^{(2)}} &= \begin{pmatrix}
        1&0&0&1
    \end{pmatrix} \begin{pmatrix}
        1 & 0 & 0 & 0\\
        0 & -1 & 0 & 0\\
        0 & 0 & 1 & 0\\
        0 & 0 & 0 & -1
    \end{pmatrix}
    \begin{pmatrix}
        1\\0\\0\\1
    \end{pmatrix} = 1-1 = 0 \\
    \ev**{\mathbb{1}_2\otimes \sigma_z^{(2)}}{1_{TM}^{(2)}} &= 
    \begin{pmatrix}
        0&1&1&0
    \end{pmatrix} \begin{pmatrix}
        1 & 0 & 0 & 0\\
        0 & -1 & 0 & 0\\
        0 & 0 & 1 & 0\\
        0 & 0 & 0 & -1
    \end{pmatrix}
    \begin{pmatrix}
        0\\1\\1\\0
    \end{pmatrix} = 1-1 = 0
    \end{cases}\right.
    \\
    &\left.\begin{cases} 
    \ev**{\sigma_z^{(1)}\otimes \sigma_z^{(2)}}{0_{TM}^{(2)}} &= \begin{pmatrix}
        1&0&0&1
    \end{pmatrix} \begin{pmatrix}
        1 & 0 & 0 & 0\\
        0 & -1 & 0 & 0\\
        0 & 0 & -1 & 0\\
        0 & 0 & 0 & 1
    \end{pmatrix}
    \begin{pmatrix}
        1\\0\\0\\1
    \end{pmatrix} = 1+1 = 2 \\
    \ev**{\sigma_z^{(1)}\otimes \sigma_z^{(2)}}{1_{TM}^{(2)}} &= 
    \begin{pmatrix}
        0&1&1&0
    \end{pmatrix} \begin{pmatrix}
        1 & 0 & 0 & 0\\
        0 & -1 & 0 & 0\\
        0 & 0 & -1 & 0\\
        0 & 0 & 0 & 1
    \end{pmatrix}
    \begin{pmatrix}
        0\\1\\1\\0
    \end{pmatrix} = -1-1 = -2
    \end{cases}\right.
    \end{aligned}$$
    The orthogonality for $N=2$ is conserved since we are dealing with diagonal operators.
    \item The orthogonality for any $N$ is trivial since the matrices are diagonal. Let's look at $\ev{\bigotimes_{k\in Z'} \sigma_z^{(k)}}{0_{TM}^{(N)}} - \ev{\bigotimes_{k\in Z'} \sigma_z^{(k)}}{1_{TM}^{(N)}}$ and use the recursive definition of the states. \\We also define $\epsilon^{(k)}=\begin{cases}
        \sigma_z^{(k)} & \text{if }\sigma_z^{(k)}\in Z'\\
        \mathbb{1}_2 & \text{if not}
    \end{cases}$.
    $$\begin{aligned}
        &\ev**{\bigotimes_{k\in Z'} \sigma_z^{(k)}}{0_{TM}^{(N)}} - \ev**{\bigotimes_{k\in Z'} \sigma_z^{(k)}}{1_{TM}^{(N)}} \\ 
        &\qquad=\ev**{\ev**{\bigotimes_{k\in Z'} \sigma_z^{(k)}}{0}}{0_{TM}^{(N-1)}} + \ev**{\ev**{\bigotimes_{k\in Z'} \sigma_z^{(k)}}{1}}{1_{TM}^{(N-1)}}\\&\qquad\quad-\ev**{\ev**{\bigotimes_{k\in Z'} \sigma_z^{(k)}}{0}}{1_{TM}^{(N-1)}} - \ev**{\ev**{\bigotimes_{k\in Z'} \sigma_z^{(k)}}{1}}{0_{TM}^{(N-1)}}
        \\ 
        &\qquad=\ev**{\bigotimes_{k\in Z'\backslash\sigma_z^{(N)}} \sigma_z^{(k)}}{0_{TM}^{(N-1)}}\ev**{\epsilon^{(N)}}{0} + \ev**{\bigotimes_{k\in Z'\backslash\sigma_z^{(N)}} \sigma_z^{(k)}}{1_{TM}^{(N-1)}}\ev**{\epsilon^{(N)}}{1}\\&\qquad\quad-\ev**{\bigotimes_{k\in Z'\backslash\sigma_z^{(N)}} \sigma_z^{(k)}}{1_{TM}^{(N-1)}}\ev**{\epsilon^{(N)}}{0} - \ev**{\bigotimes_{k\in Z'\backslash\sigma_z^{(N)}} \sigma_z^{(k)}}{0_{TM}^{(N-1)}}\ev**{\epsilon^{(N)}}{1}
        \\
        &\qquad=\left(\ev**{\bigotimes_{k\in Z'\backslash\sigma_z^{(N)}} \sigma_z^{(k)}}{0_{TM}^{(N-1)}}-\ev**{\bigotimes_{k\in Z'\backslash\sigma_z^{(N)}} \sigma_z^{(k)}}{1_{TM}^{(N-1)}}\right)\ev**{\epsilon^{(N)}}{0} \\&\qquad\quad+ \left(\ev**{\bigotimes_{k\in Z'\backslash\sigma_z^{(N)}} \sigma_z^{(k)}}{1_{TM}^{(N-1)}} - \ev**{\bigotimes_{k\in Z'\backslash\sigma_z^{(N)}} \sigma_z^{(k)}}{0_{TM}^{(N-1)}}\right)\ev**{\epsilon^{(N)}}{1}
    \end{aligned}$$
    If we assume the property holds for $N-1$, this expression equals to zero if
    \begin{itemize}
        \item $\sigma_z\in Z'$ and $Z'\backslash\sigma_z^{(N)}$ doesn't already contain all the other $N-1$ $\sigma_z^{(k)}$s.\\ Since $\sigma_z\in Z'$, this means $\epsilon^{(N)} = \sigma_z$ which implies $\ev{\epsilon^{(N)}}{0} = -\ev{\epsilon^{(N)}}{1}$, so in order for the difference to be zero, we need $\ev{\bigotimes_{k\in Z'\backslash\sigma_z^{(N)}} \sigma_z^{(k)}}{0_{TM}^{(N-1)}}-\ev{\bigotimes_{k\in Z'\backslash\sigma_z^{(N)}} \sigma_z^{(k)}}{1_{TM}^{(N-1)}} = 0$, which does hold true as long as $Z'\backslash\sigma_z^{(N)}$ doesn't already contain all the other $N-1$ $\sigma_z^{(k)}$s.
    \end{itemize}
     or 
     \begin{itemize}
         \item $\sigma_z \notin Z'$ since $\epsilon^{(N)} = \mathbb{1}_2 \Rightarrow \ev{\epsilon^{(N)}}{0} = \ev{\epsilon^{(N)}}{1}$. 
     \end{itemize}
     So, $ \ev{\bigotimes_{k\in Z'} \sigma_z^{(k)}}{0_{TM}^{(N)}} - \ev{\bigotimes_{k\in Z'} \sigma_z^{(k)}}{1_{TM}^{(N)}} = 0$ iff $E'$ doesn't contain all $\sigma_z^{(k)}s.$

     A similar argument can be made with $\sigma_y = -i\sigma_z\sigma_x$ using the diagonality of $\sigma_z$.
\end{itemize}

\subsection{Proof of Property 1.2.3}

\textit{$\forall j < N, i\in\{y,z\}$ \todo*{check for $\sigma_y$} $$\ev**{\left(\sum_{k=1}^N \sigma_{i}^{(k)}\right)^{j}}{0_\text{TM}^{(N)}} = \ev**{\left(\sum_{k=1}^N \sigma_{i}^{(k)}\right)^{j}}{1_\text{TM}^{(N)}},$$ $$\text{and }\mel**{1_\text{TM}^{(N)}}{\left(\sum_{k=1}^N \sigma_{i}^{(k)}\right)^{j}}{0_\text{TM}^{(N)}} = 0.$$}

For a given integer $j$, $$\left(\sum_{k=1}^N \sigma_{y,z}^{(k)}\right)^{j} =  \sum_{j_1+j_2+\cdots+j_N=j}\frac{j!}{j_1!\, j_2! \cdots j_m!}
  \prod_{k=1}^N \left(\sigma_{y,z}^{(k)}\right)^{j_k},$$
which is then a state-independent weighted sum of products over ensembles $E$ like in property 2. The ensembles $E$ are then simply partitions of $j$ of size $N$, and as long as $j<N$, it is not possible for $E$ to contain all $\sigma_z^{(k)}$s. If $j\geq N$, there exists ensembles $E$ such that the equalities of property 1.2.2 do not hold anymore. Therefore property 1.2.3 is a consequence of property 1.2.2.

\subsubsection{qudit version for $J_z$}

If one works with PTM logical states in a qudit of dimension $d=2^N$.
\begin{equation*}
     \ket{1_\text{TM}^{(N)}} = \sqrt{\frac{2}{d}} \sum_{k=0}^{d-1} t_k \ket{k} = \sqrt{\frac{2}{d}} \sum_{k\in O(N)} \ket{k} \;\text{ and } \;
     \ket{0_\text{TM}^{(N)}} = \sqrt{\frac{2}{d}} \sum_{k=0}^{d-1} \bar{t_k} \ket{k} = \sqrt{\frac{2}{d}} \sum_{k\in E(N)} \ket{k} 
\end{equation*}

We have that 
$$\begin{aligned}
    \ev**{J_{z}^{j}}{0_\text{TM}^{(N)}} &= \sum_{k=0}^{d-1} \left(\frac{d-1-2k}{2}\right)^j\\
    &= 2^{-j} \sum_{k\in E(N)} \sum_{l=0}^j \begin{pmatrix}j\\n\end{pmatrix} 2^l k^l (d-1)^{j-l} \\
    &= 2^{-j} \sum_{l=0}^j \begin{pmatrix}j\\n\end{pmatrix} 2^l \left(\sum_{k\in E(N)} k^l\right) (d-1)^{j-l}, 
\end{aligned}$$
and similarly,
$$\ev**{J_{z}^{j}}{1_\text{TM}^{(N)}} = 2^{-j} \sum_{l=0}^j \begin{pmatrix}j\\n\end{pmatrix} 2^l \left(\sum_{k\in O(N)} k^l\right) (d-1)^{j-l}.$$

Refering to the discussion about the Prouhet-Tarry-Escott problem, we know that $\sum_{k\in E(N)} k^l = \sum_{k\in O(N)} k^l$, and, therefore $\ev**{J_{z}^{j}}{0_\text{TM}^{(N)}}=\ev**{J_{z}^{j}}{1_\text{TM}^{(N)}}$. Moreover, $J_z$ is diagonal therefore $\mel**{1_\text{TM}^{(N)}}{J_z^{j}}{0_\text{TM}^{(N)}} = 0$.

\subsection{Proof of Property 1.2.4}

$$\forall k,j < N,\quad \sigma_x^{(k)}\sigma_x^{(j)}\ket{0_\text{TM}^{(N)}} = \ket{0_\text{TM}^{(N)}} \text{ and }\sigma_x^{(k)}\sigma_x^{(j)}\ket{1_\text{TM}^{(N)}} = \ket{1_\text{TM}^{(N)}} $$

Since \(\sigma_x^2 = \mathbb{1}\), if \(k = j\), then the state remains unchanged. If exactly two different qubits are flipped without a relative phase (which is the action of two \(\sigma_x\) operators), the parity of the number of 1's stays unchanged.

\end{document}